%% file: main.tex
\documentclass[letterpaper,11pt]{article}

\usepackage[utf8]{inputenc}
\usepackage[letterpaper,left=2cm,right=2cm,top=2cm,bottom=2cm]{geometry}
\usepackage[justification=centering]{subfig}
\usepackage{authblk}
\usepackage{url}

\usepackage{amsmath,graphicx,amssymb}
\usepackage{dirtytalk}
\usepackage{color}
\usepackage{float}

\usepackage{xcolor}

\input{math_commands.tex}

\title{Normalizing flows for random fields in cosmology}
\author[1]{Adam Rouhiainen\thanks{Electronic address: rouhiainen@wisc.edu}}
\author[2]{Utkarsh Giri}
\author[1]{Moritz M\"{u}nchmeyer}

\affil[1]{Department of Physics, University of Wisconsin-Madison, Madison, WI 53706, USA}
\affil[2]{Perimeter Institute for Theoretical Physics, Waterloo, ON N2L 2Y5, Canada}

\begin{document}

\maketitle

\begin{abstract}
Normalizing flows are a powerful tool to create flexible probability distributions with a wide range of potential applications in cosmology. Here we are studying normalizing flows which represent cosmological observables at field level, rather than at the level of summary statistics such as the power spectrum. We evaluate the performance of different normalizing flows for both density estimation and sampling of near-Gaussian random fields, and check the quality of samples with different statistics such as power spectrum and bispectrum estimators. We explore aspects of these flows that are specific to cosmology, such as flowing from a physical prior distribution and evaluating the density estimation results in the analytically tractable correlated Gaussian case. 
\end{abstract}

\section{Introduction and brief review of normalizing flows}

Normalizing flows are a major recent development in probabilistic machine learning (see \cite{2019arXiv191202762P} for a review) because they provide a powerful tool to represent flexible and expressive probability densities. A flurry of recent work has shown their usefulness for a wide range of applications, such as image generation \cite{2018arXiv180703039K,2016arXiv160508803D}, variational inference \cite{rezende2016variational}, and likelihood-free inference \cite{2017arXiv170507057P,2019arXiv191101429C}. In cosmology, normalizing flows have recently been used to represent the posterior distribution of summary statistics such as the power spectrum or cosmological parameters \cite{Alsing:2019xrx,DiazRivero:2020oai}. In the present paper we use normalizing flows to represent the probability density of fields such as the matter distribution directly at field level.

We start with a brief review of normalizing flows. A normalizing flow is a natural way to construct flexible probability distributions by transforming a simple base distribution (often Gaussian, thus the name) into a complicated target distribution. This is done by applying a series of learned diffeomorphisms to the base distribution. Given a base distribution $p_u(\uvec)$ of a random variable $\uvec$, the target distribution $p_x(\xvec)$ is given by
\begin{equation}
    p_x(\rvx) = \pu(\rvu)\,\abs{\det J_T(\rvu)}^{-1}
\end{equation}
where $T$ is the transformation (the ``flow") $\rvx = T(\rvu)$ and $J_T$ is its Jacobian. We can construct a transformation $T$ by composing a finite number of simple transformations $T_k$ as follows:
\begin{equation}
    T = T_{K} \circ \cdots \circ T_{1}.
\end{equation}
These simple transformations must have the property that they have tractable inverses and a tractable Jacobian determinant. They depend on learned parameters and can be parametrized using neural networks. In this way very expressive densities can be constructed. Assuming $\rvz_0=\rvu$ and $\rvz_K=\rvx$, the transformation at each step $k$ is
\begin{align}
    \rvz_{k} = T_k(\rvz_{k-1})
\end{align}
and the Jacobian determinant is:
\begin{equation}
\log \left| J_{T}(\rvz)  \right| = \sum_{k=1}^{K} \log \left| J_{T_{k}}(\rvz_{k-1})  \right|.
\end{equation}

Once the flow is learned, two basic statistical operations can be performed efficiently: density evaluation (what is $p_x(\xvec)$ given a sample $\xvec$) and sampling from $p(\xvec)$. These operations can be used for statistical inference purposes. The difference between normalizing flows and ordinary neural network techniques are that the former are representing normalized probability densities rather than arbitrary mappings from input to output. A number of flow architectures that are applicable to 2D fields (images) have been proposed \cite{2019arXiv191202762P}, some of which can generate very high quality images. While normalizing flows are often not quite as expressive as Generative Adverserial Networks (GANs) or Variational Autoencoders (VAE), they come with the statistical interpretation of a probability density function (PDF) and in particular offer exact density evaluation. 

It is a priori very plausible that normalizing flows could be strong at representing PDFs of fields in cosmology. Many fields in cosmology, such as the matter distribution, start as Gaussian fields in the far past. They then become progressively more non-Gaussian with time due to non-linear interactions. In the same way, a Gaussian field base distribution of a normalizing flow becomes progressively more non-Gaussian by the applications of more transformations $T_k$. Such a normalizing flow is therefore a natural candidate to represent somewhat (non-)Gaussian PDFs of matter fields at late times.

Normalizing flows for fields in cosmology can be used for data analysis in several different ways. In particular we will pursue the following applications in the future:
\begin{itemize}
    \item Representing non-Gaussian prior distributions of small-scale fields in cosmology for better posterior analyses. For example, the small-scale Cosmic Microwave Background (CMB) lensing potential (see e.g.~\cite{Lewis:2006fu}) is not Gaussian and using a non-Gaussian prior $\mathcal{P}(\phi)$, learned from simulations, in a MAP or HMC analysis would allow to increase the signal-to-noise under suitable conditions.
    \item Variational inference at field level. Variational inference (see e.g.~\cite{2016arXiv160100670B}) fits an approximate posterior distribution to samples from the unknown true posterior. It is usually more efficient than Monte Carlo methods. Normalizing flows are ideally suited for variational inference \cite{rezende2016variational}. In our lensing potential example, this would mean to make a variational estimate of the lensing potential rather than a MAP as in \cite{Carron:2017mqf} or an HMC sampling as in \cite{Millea:2020iuw}. Hamiltonial Variational Inference \cite{2014arXiv1410.6460S} may also be useful for this purpose.
\end{itemize}
We have used CMB lensing as an example here but there are many other analyses in cosmology that are mathematically similar. In addition, the trained flow can also be used in forward direction for sampling, for example to sample from foreground contamination in CMB data analysis, as recently shown in \cite{Thorne:2021nux} using variational auto encoders rather than normalizing flows. Before we can use normalizing flows for these challenging practical applications it is important to quantify their performance on simpler problems. This is the purpose of this study.

\section{Flow architecture and training process}

In this section we will review the architecture of the normalizing flow, i.e.\ the form of the transformations $T_k$ and prior $p_u$. We start with a basic flow called real NVP, which we use in the main text of this paper, and then briefly discuss some generalisations, one of which we apply to the data in App.~\ref{sec:glow}. In general, the goal of constructing normalizing flows is to make them both flexible as well as computationally tractable, and the flows discussed here have both of these properties. For a more detailed review of the motivation behind these constructions we refer to the original papers.

\subsection{Basic real NVP flow}
\label{sec:realnvp}

The first flow with success in creating high quality images was the real-valued non-volume preserving (\emph{real NVP}) flow \cite{2016arXiv160508803D}. This flow is expressive and is fast both for sampling and inference. This flow has also recently received a lot of attention for its use to represent PDFs in lattice field theory (see e.g.~\cite{Albergo:2021vyo}). This is the basic flow we will be using in this study, but we will consider a more complicated flow in App.~\ref{sec:glow}. A simplification of our application compared to real NVP is that the latter was constructed for RGB images (3 channels) while we represent a scalar field (1 channel), so we do not require transformations that mix channels.

\textbf{Affine coupling layer}. The basis of this flow is the \textit{affine coupling layer}, an operation that rescales and shifts a subset of the random variables depending on the value of the other random variables. Split the set of all random variables (in our application all pixels of the random field), denoted as $x$, into two subsets $x_1$ and $x_2$. Then update these parameters as follows:
\begin{align}
    x_1' &= e^{s(x_2)} x_1 + t(x_2) \\
    x_2' &= x_2
\end{align}
Here $s$ and $t$ are vector valued, i.e.\ each pixel in $x_1$ can be rescaled and shifted differently. This transformation guarantees invertibility as well as a triangular Jacobian which is computationally very easy to evaluate and invert. While the scaling transformation is simple, its flexibility comes from the free form of the functions $s$ and $t$ and the fact that we stack many such layers, with different partitions into subsets.

\textbf{Checkerboard masking}. For images, an elegant choice of partition into subsets $x_1$ and $x_2$ is the checkerboard masking proposed in \cite{2016arXiv160508803D}. On the checkerboard, we either use the \say{white} pixels for $x_1$ and the \say{black} pixels for $x_2$ or vice versa. For each consecutive affine coupling layer we switch the sets $x_1$ and $x_2$. 

\textbf{CNN to determine the affine parameters}. We now need to define the functions $s(x_2)$ and $t(x_2)$. As $x_2$ are spatially organized, rather than arbitrary collections of random variables, it is natural to use a standard convolutional neural network for this purpose (as was done in \cite{2016arXiv160508803D}) which enforces translational symmetry. The CNN needs to conserve dimension, so we use stride 1 convolutions and no pooling. To implement periodic boundary conditions we use the common approach of circular padding. As in \cite{Albergo:2021vyo} we use 3 convolutional layers with kernel size 3 and leaky ReLU activation functions. The number of channels is in this order: 1 (the scalar PDF values), 16 (arbitrary number of feature maps), 16 (arbitrary number of feature maps), 2 (the output variables $s$ and $t$).

\textbf{Stacking the layers}. As our default configuration, as in \cite{Albergo:2021vyo}, we stack $K=16$ affine coupling layers, each with their own CNN to parametrise the affine transformation $s$ and $t$.

This architecture provides a total of 44,320 trainable parameters. Our PyTorch \cite{NEURIPS2019_9015} implementation of this normalizing flow is taken from \cite{Albergo:2021vyo}, with minor modifications such as a different loss function and different priors. An illustration of the architecture can be found in Fig.~\ref{fig:architectures}(a) in App.~\ref{sec:glow}.

\subsection{Other flows of interest}
There are a number of interesting generalisations of the real NVP flow:
\begin{itemize}
    \item \textbf{Real NVP with multi-scale architecture}. The original real NVP paper \cite{2016arXiv160508803D} used a multi-scale architecture to help with representing larger images. The idea is to reshape the random field to a smaller spatial dimension while increasing the number of channels, thus conserving the number of variables as required. One can then mix the channels in different ways. 
    \item \textbf{Glow}. A well-known extension of the real NVP flow with multi-scale architecture is the Glow \cite{2018arXiv180703039K} flow, which generates very high quality images. Glow mixes channels with $1\times1$ convolutions and does not use a checkerboard to split spatial dimension.
    \item \textbf{Periodic convolutional flows}. Another natural flow architecture which is applicable to our case is the periodic convolutional flow proposed in \cite{2019arXiv190111137H}. This flow makes use of the fact that convolutions on periodic spaces can be inverted efficiently in Fourier space due to the convolution theorem. The applications below can be made periodic at least with sufficient zero padding. 
\end{itemize}
We provide results for the Glow flow in App.~\ref{sec:glow} and defer investigation of convolutional flow architectures to future work.

\subsection{Flow training}

To train the flow, we minimize the forward Kullback–Leibler  (KL) divergence (see Sec.~2.3 of the review \cite{2019arXiv191202762P}), a measure of the relative entropy from the target distribution $p^*_x(\rvx)$ to the base distribution $p_x(\rvx)$. The forward KL divergence can be expressed as
\begin{align}
    \mathcal{L(\phi)}&=D_{\text{KL}}\big(p^*_x(\rvx)\ ||\ p_x\left(\rvx;\phi\right)\big)\\
    &=-\mathbb{E}_{p^*_x(\rvx)}\big(\log p_x(\rvx;\phi)-\log p^*_x(\rvx;\phi)\big)\\
    &=-\mathbb{E}_{p^*_x(\rvx)}\big(\log\pu\left(T^{-1}\left(\rvx;\phi\right)\right)+\log\left|\det{J_{T^{-1}}(\rvx;\phi)}\right|\big)+\mathbb{E}_{p^*_x(\rvx)}\log p^*_x(\rvx;\phi)
\end{align}
where $T$ is the flow transformation, with learned parameters $\phi$. The final term here is a constant that we do not need to calculate. Here \say{forward} denotes the order of $p^*_x$ and $p_x$ above. The expectation values are estimated as
\begin{equation}\label{eq:loss}
    \mathcal{L}\left(\phi\right)=-\frac{1}{N}\sum_{n=1}^N\left(\log\pu\left(T^{-1}\left(\rvx_n;\phi\right)\right)+\log\left|\det J_{T^{-1}}(\rvx_n;\phi)  \right|\right)+\text{const.},
\end{equation}
where $\rvx_n$ are the training samples from $p^*_x(\rvx)$. Minimizing the Monte Carlo approximation of the KL divergence is thus equivalent to fitting the flow model to the training samples by maximum likelihood estimation. We use the Adam optimizer to minimize the loss with respect to the parameters $\phi$. Unless otherwise specified we use a learning rate of $0.001$ and batch size of 128. We also normalize the training samples to unit variance.

\section{Application to Gaussian fields}

As the first toy application, we show that we can flow from random uncorrelated Gaussian noise to a correlated Gaussian random field with a known power spectrum. We verify both the quality of the samples, as well as the quality of the density estimation operation. While in the Gaussian case we can sample efficiently and evaluate the density analytically and the flow treatment has thus no practical value, it is the basis for extensions to intractable non-Gaussian fields.

\subsection{Sample quality}

\begin{figure}[h!]
\centering
\includegraphics[width=0.8\textwidth]{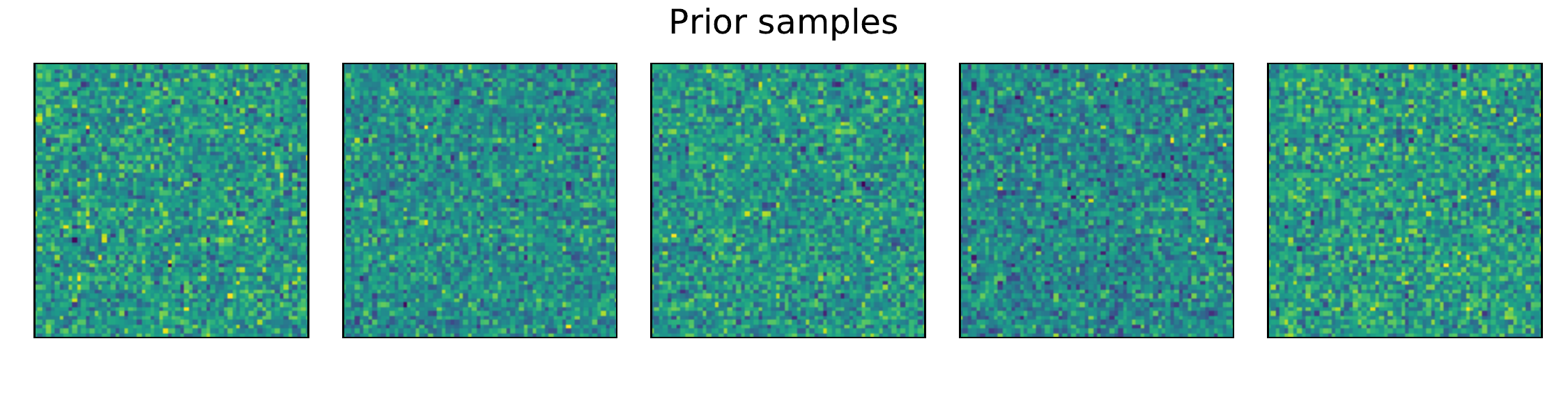}\\
\includegraphics[width=0.8\textwidth]{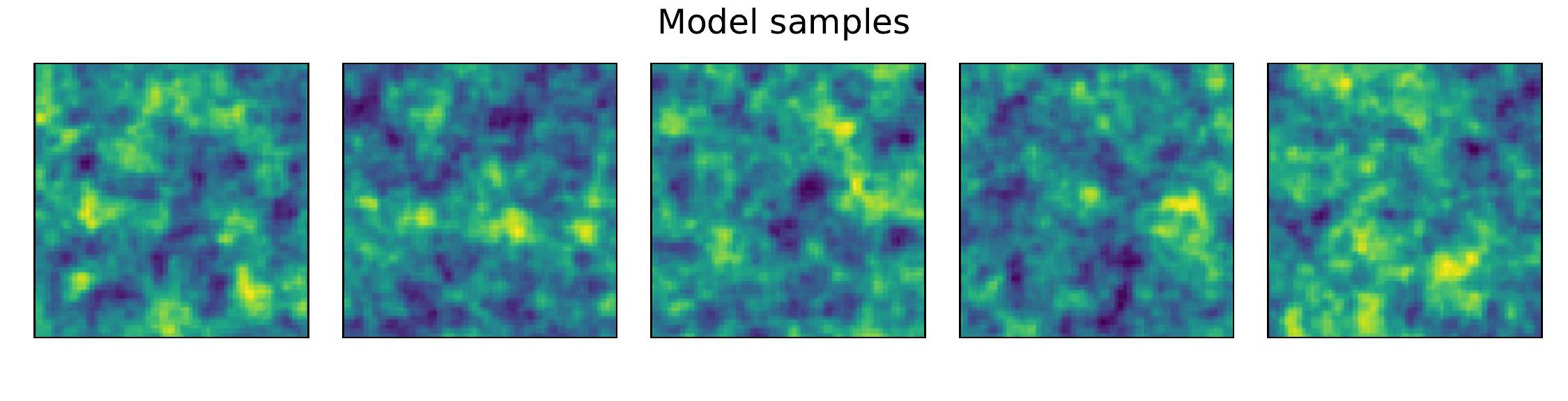}\\
\includegraphics[width=0.8\textwidth]{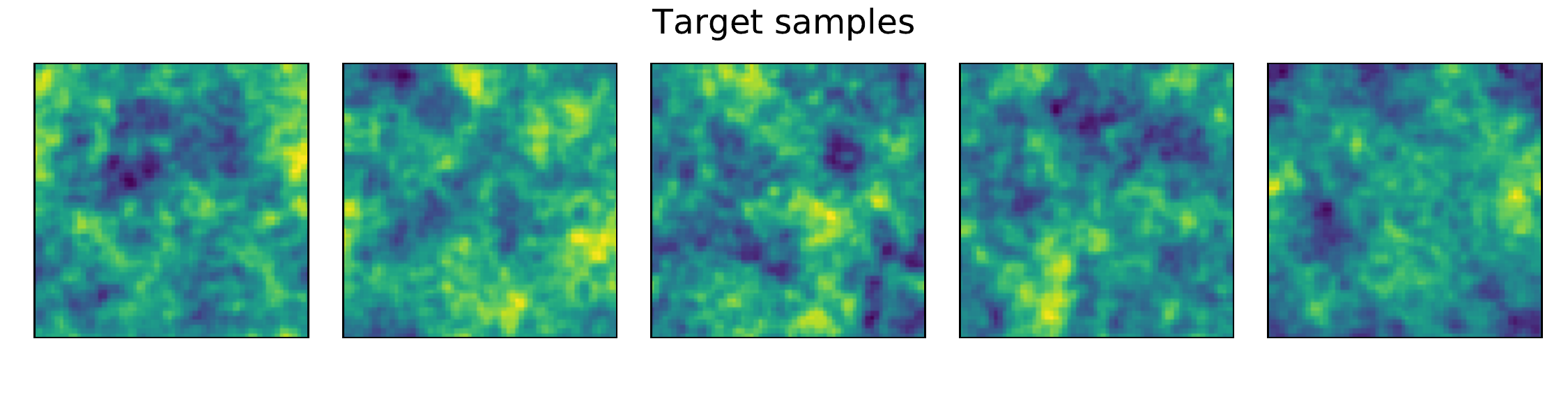}
\caption{Gaussian field samples with a CMB power spectrum. Random samples from the prior (top), corresponding flow (middle), and training (bottom). The training data is using a CMB power spectrum on a 4 degree sky patch on $64\times64$ pixels.}\label{fig:gausssamples1}
\end{figure}

\begin{figure}[h!]
\centering
\includegraphics[width=0.45\textwidth]{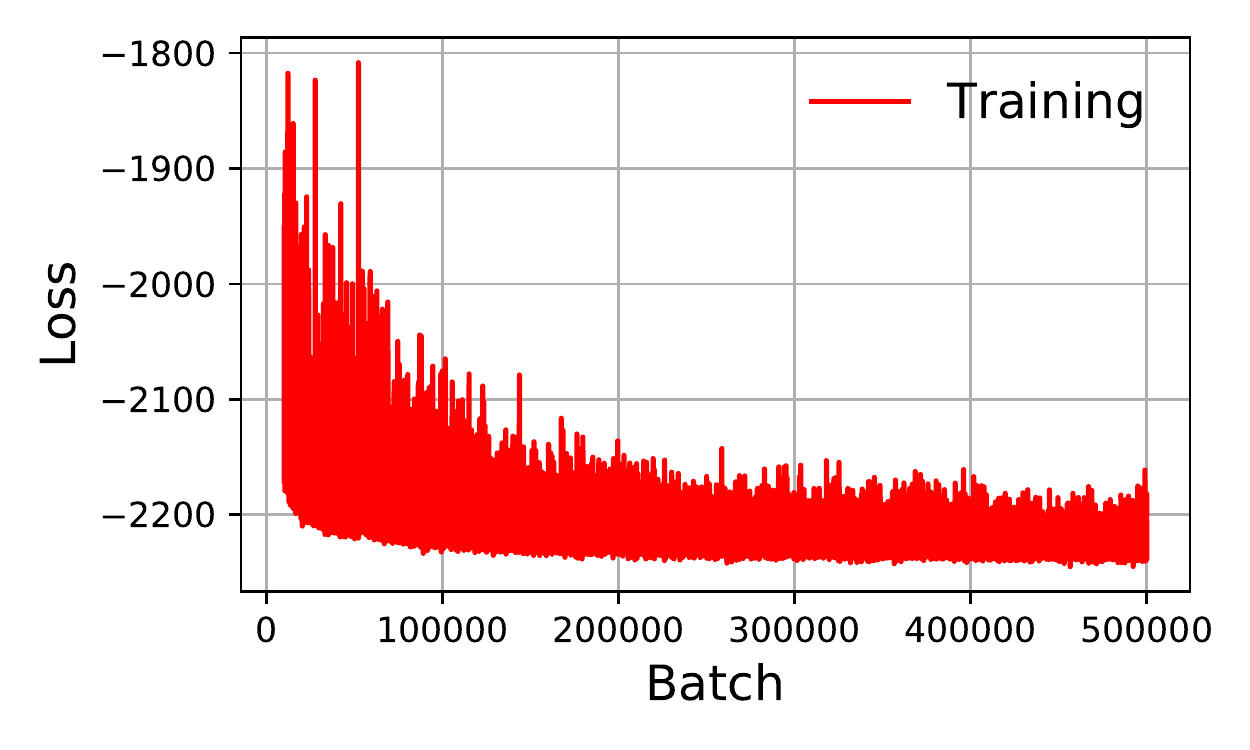}
\includegraphics[width=0.45\textwidth]{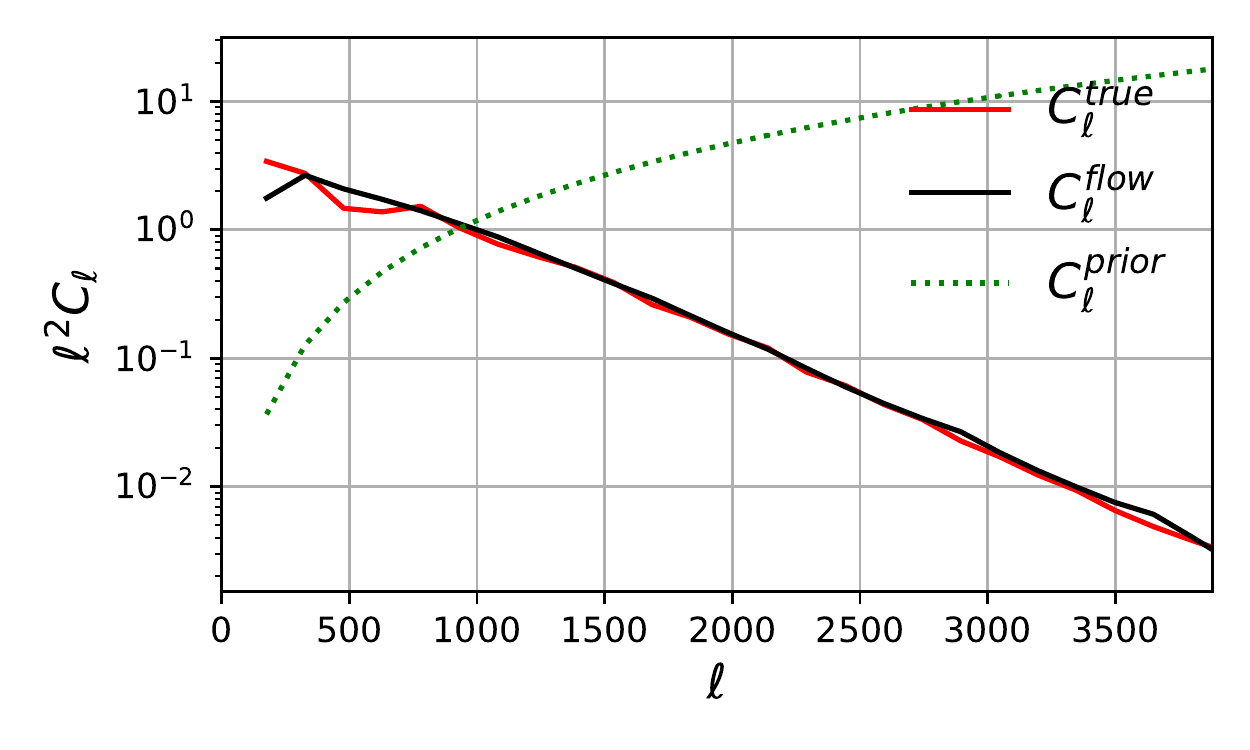}\\
\includegraphics[width=0.45\textwidth]{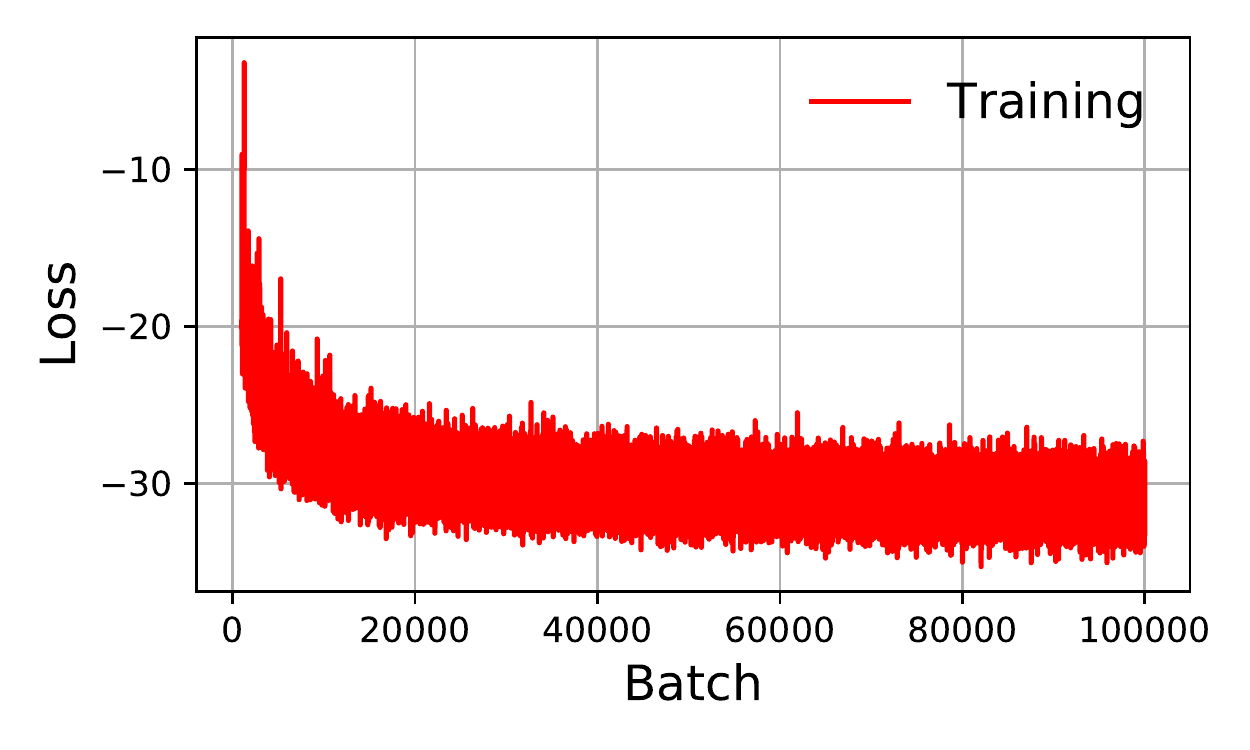}
\includegraphics[width=0.45\textwidth]{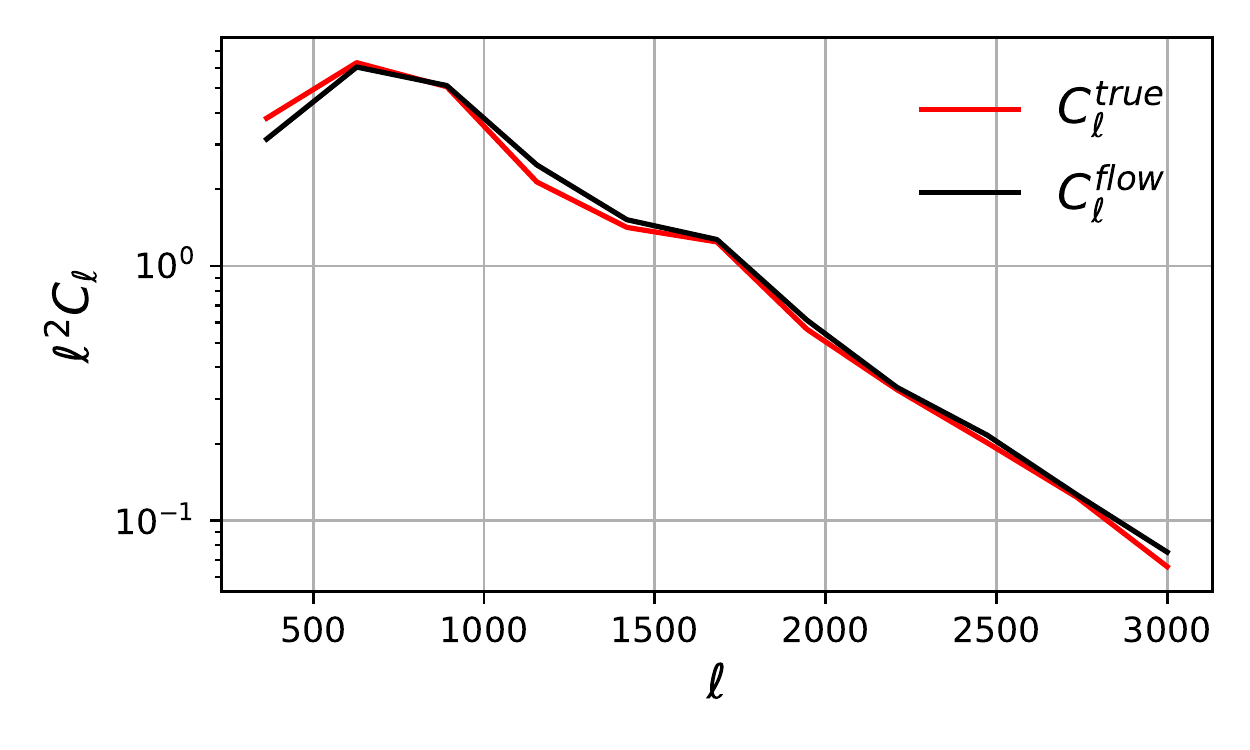}\\
\caption{Top left: Loss convergence of the 64px side length Gaussian field training (starting from batch 10,000). Top right: Comparison of the prior, model, and training set power spectrum, averaged over 10,000 maps. The x-axis is the multipole expansion coefficient $\ell$. While the model power spectrum is generally correct, the 64px case (top row) does not resolve the BAO wiggle. On the bottom we show the same plots for 16px maps. Here the BAO is resolved better. We improve BAO and training convergence results below.}\label{fig:gaussps}
\end{figure}

We train the real NVP flow described in Sec.~\ref{sec:realnvp} on samples from a correlated Gaussian field, with a CMB temperature power spectrum. We use this as an example power spectrum from cosmology with baryon acoustic oscillations (BAO) features in it, but we are not interested in representing Gaussian primary CMB for practical applications. Here we flow from an uncorrelated Gaussian prior distribution. We use patches of $64\times64$ pixels covering a sky angle side length of 4 degrees. We have experimented with infinite \say{on the fly} created training samples as well as fixed size training samples with 10,000 or 1,000 samples. We show the infinite case here, but we only find a difference with limited training data for the density estimation task which we discuss below. Samples from the prior, model, and target (training) distribution are shown in Fig.~\ref{fig:gausssamples1}. By eye, the model samples look like the training data. We also compare their power spectrum in Fig.~\ref{fig:gaussps} and find good agreement. The power spectrum has however not fully resolved the BAO feature. It is likely that with longer training and more network capacity this could be improved. However, we will fix this problem in the next section more elegantly using a correlated prior, and in App.~\ref{sec:glow} using a different flow architecture. The total training time was about 40 hours on an RTX 3090 for the 64px maps, and uses 3.8 GB of GPU memory. We will also find much faster convergence with a correlated prior in the next section.

\subsection{Density evaluation}

For some of the data analysis applications discussed in the introduction, we need to use the flow in reverse direction for density evaluation. There are two different tasks we can consider here: in distribution density evaluation (IID: independent, identical distribution) and out of distribution (OOD) density evaluation. We focus here on IID density evaluation, leaving OOD to future work. 

\begin{figure}[h!]
\centering
\includegraphics[width=0.9\textwidth]{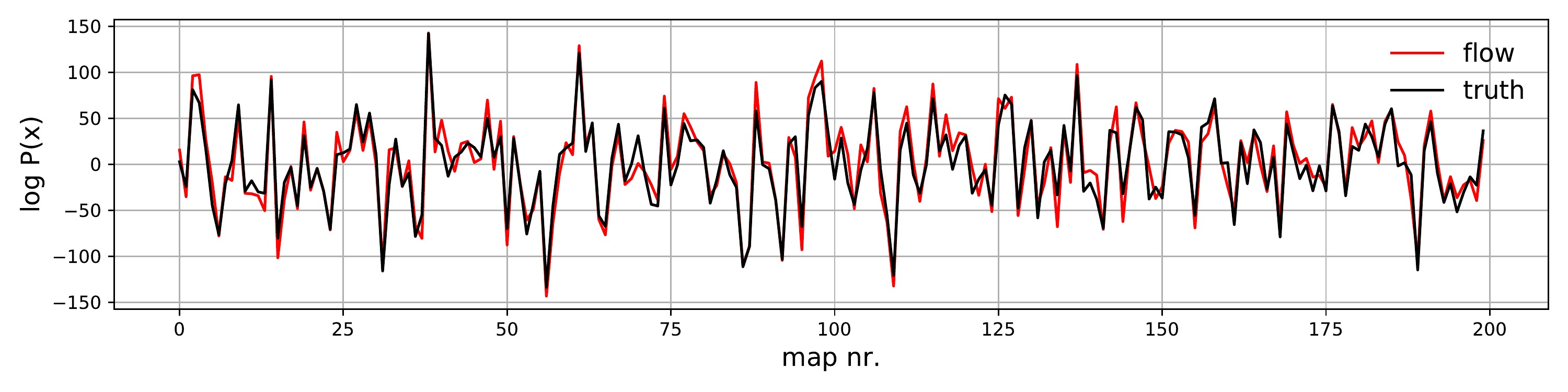}
\caption{Density evaluation $\log P(x)$ result from the flow compared to the true log probability of the sample (mean shifted to zero), for 200 maps with $64\times64$ pixel resolution. The cross-correlation coefficient is 0.993 (estimated from 10,000 maps). The flow has thus learned density evaluation on new IID samples. Here we used on the fly created training data, see main text for results with limited training data.}
\label{fig:density_estimate_gauss}
\end{figure}

We would like to verify that IID samples, when run backwards through the flow (with uncorrelated Gaussian noise prior), are assigned a probability that corresponds to their true probability. Here the fact that we start with a tractable Gaussian distribution allows us to compare the flow probability with the exact analytic probability. We sample 10,000 new IID samples $x$ from the same distribution as the training data and reverse flow them to obtain $\log P^{\mathrm{flow}}(x)$. For the configuration above, we find that the cross correlation coefficient between $\log P^{\mathrm{flow}}(x)$ and $\log P^{\mathrm{true}}(x)$ is $r \simeq 0.993$. We plot these quantities for 200 random example maps in Fig.~\ref{fig:density_estimate_gauss}. We thus find that the flow can be used for density evaluation in reverse mode with high accuracy.

We examined whether precise density evaluation results were only possible due to the on the fly created training data, which ensured that the training data covered a large number of independent points in sample space. Indeed we found that by training on 1,000 Gaussian maps, the cross correlation coefficient dropped to about $r \simeq 0.97$, while with 10,000 Gaussian maps we found $r \simeq 0.98$. At the level of the power spectrum we found no difference with respect to the on the fly created training data, and we did not reach a domain of over-training with the limited training data. Generating non-Gaussian training data can be computationally expensive in practice. Depending on the size of the training data set, different sample augmentation techniques could be used to get the best possible density evaluation results after training.

\section{Local non-Gaussianity and correlated prior}

After verifying the flow performance in the simple case of an analytically tractable correlated Gaussian field, we now consider perhaps the simplest non-Gaussian random field in cosmology, the so-called \say{local non-Gaussianity}. Local non-Gaussianity \cite{Komatsu:2001rj} is generated by transforming a correlated Gaussian field $\phi_{G}(\bx)$ as
\begin{align}
\label{eq:fnlmapmake}
 \phi_{NG}(\bx) = \phi_{G}(\bx) + \tilde{f}_{NL}^{\mathrm{local}}\left( \phi_{G}(\bx)^2 - \< \phi_{G}(\bx)^2 \>\right)
\end{align}
It is easy to draw samples from this distribution, since all we need to do is to square the Gaussian field and add it to the original field with some amplitude. This form of non-Gaussianity is generated in cosmology for example by multi-field inflation (see e.g.~\cite{Byrnes:2010em}). We will discuss a second form of non-Gaussianity in App.~\ref{sec:nongaussestim}. Unlike in cosmology, here we normalize the random field so that $\phi_{G}$ has variance 1, so that $\tilde{f}_{NL}^{\mathrm{local}} = 1$ indicates $\mathcal{O}(1)$ non-Gaussianity. To make this difference clear we have added the tilde to the notation. We are primarily interested in strongly non-Gaussian fields here, induced by gravitation and astrophysics, not in the tiny effects generated during inflation.

\begin{figure}[h!]
\centering
\includegraphics[width=0.8\textwidth]{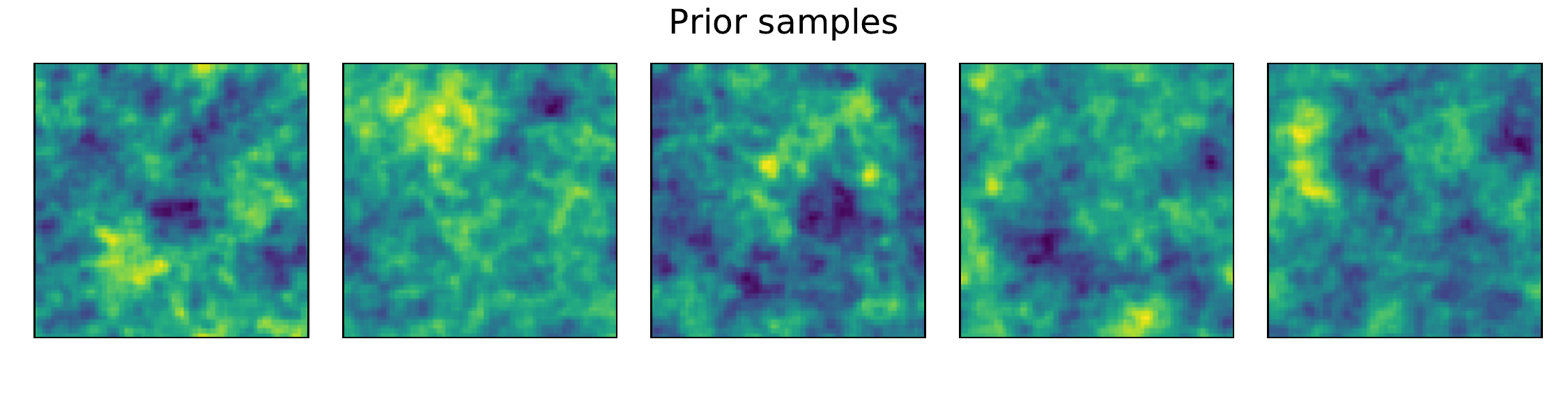}\\
\includegraphics[width=0.8\textwidth]{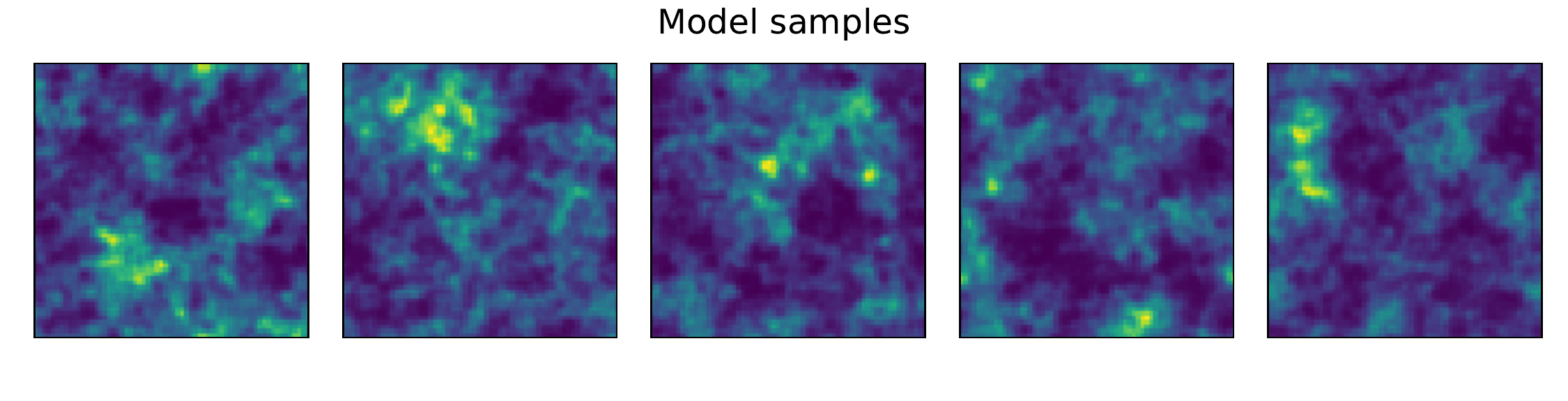}\\
\includegraphics[width=0.8\textwidth]{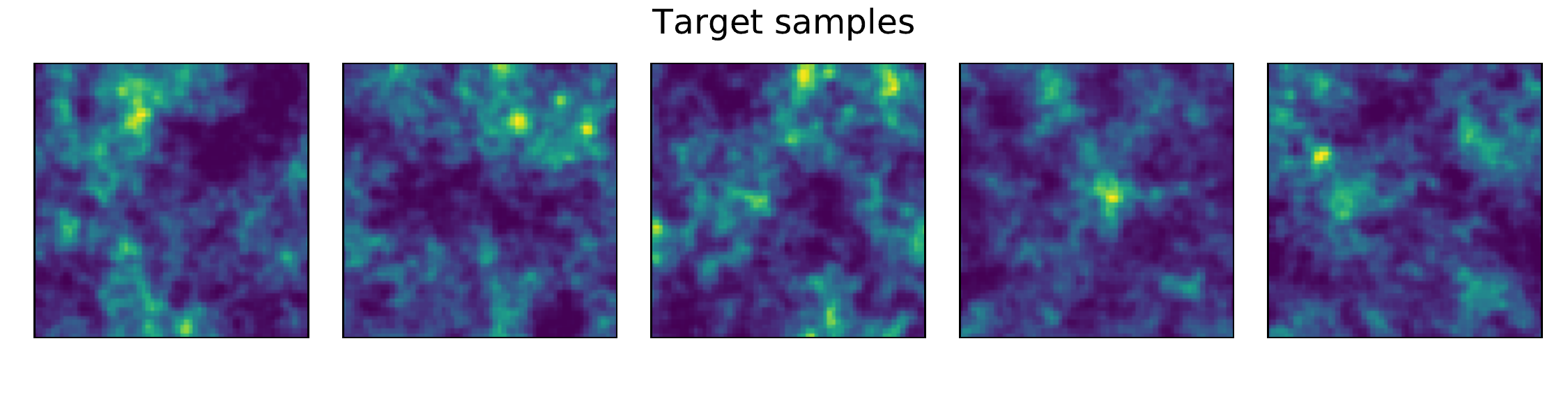}
\caption{Local non-Gaussianity field samples with $\tilde{f}_{NL}^{\mathrm{local}} = 0.2$. Gaussian prior samples (top), flow samples (middle), and training samples (bottom). The training data is generated by drawing Gaussian maps from a CMB power spectrum on a 4 degree sky patch represented on $64\times64$ pixels and making them non-Gaussian with Eq.~\ref{eq:fnlmapmake}. Comparing the prior and model samples, we find that the network makes the Gaussian prior samples more non-Gaussian, with maxima and minima becoming more pronounced.}
\label{fig:fnlsamples1}
\end{figure}

\begin{figure}[h!]
\centering
\includegraphics[width=0.45\textwidth]{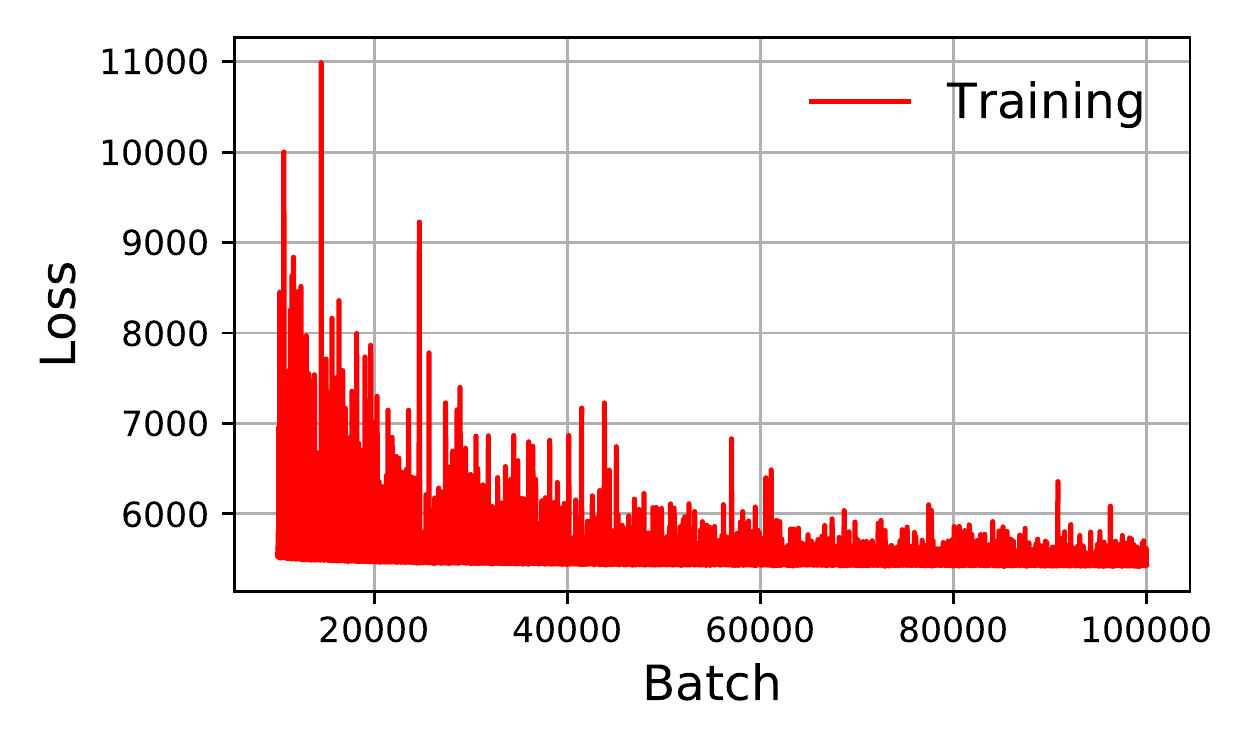}
\includegraphics[width=0.45\textwidth]{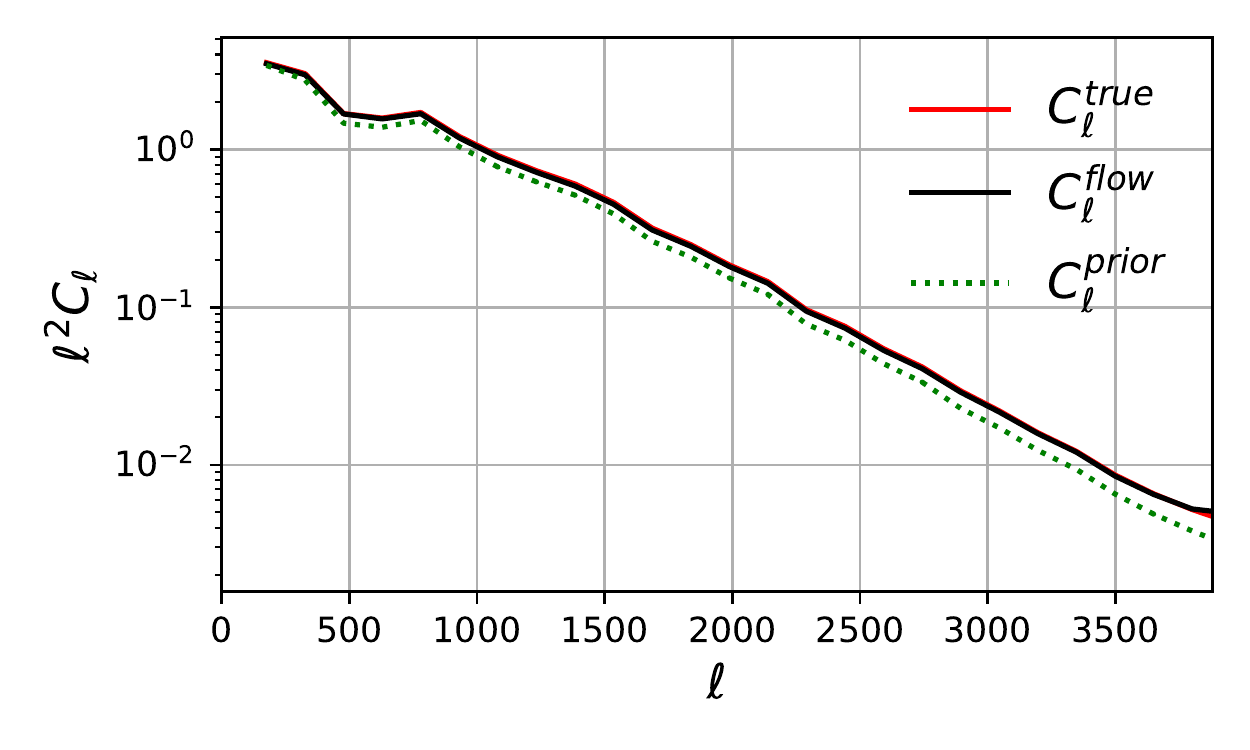}\\
\caption{Left: Loss during training (starting from batch 10,000) for the  $\fnl=0.2$ example with correlated prior. Convergence is about 10 times faster than with the random uncorrelated prior. Right: Comparison of the prior, model, and training set power spectrum averaged over 10,000 maps, for the $\fnl=0.2$ example. With the correlated prior, the BAO wiggle is resolved perfectly.}\label{fig:fnlps}
\end{figure}

We train precisely the same flow model as in the previous section, with the same training parameters. However, we now flow from a 
\emph{correlated Gaussian field with the correct power spectrum}. This makes the training much more efficient as the flow has a better starting point. In practice for the non-Gaussian fields of interest such as the matter field, it always makes sense to flow from a Gaussian field with the right power spectrum. We show samples from the prior, model, and training distribution in Fig.~\ref{fig:fnlsamples1}. As for the Gaussian case, we find that the samples are indistinguishable by eye from the training data. We show the power spectrum of the samples in Fig.~\ref{fig:fnlps} (right), including the power spectrum of the prior that now matches the training data. The flow thus learns to induce the right non-Gaussianity while keeping the power spectrum intact. The correlated prior also improves the training convergence. The flow trains roughly 10 times faster than with the uncorrelated prior, as illustrated by the training loss in Fig.~\ref{fig:fnlps} (left).

We now measure the non-Gaussianity in the samples with the method reviewed in App.~\ref{sec:nongaussestim}. We estimate the amplitude of two non-Gaussian templates, the so-called local and equilateral non-Gaussianity. In the present case the training data has local non-Gaussianity by definition, however the equilateral template has some overlap with the local template, and it is useful to measure both in the model samples whose non-Gaussianity can be non-local. We find the following mean estimated values, averaging over 10,000 training and flow samples with 64px resolution:

\begin{center}
\begin{tabular}{ |c||c|c|c| } 
 \hline
 $\tilde{f}_{NL}^{\mathrm{local}}$ training & $\tilde{f}_{NL}^{\mathrm{local}}$ flow & $\tilde{f}_{NL}^{\mathrm{equi}}$ training & $\tilde{f}_{NL}^{\mathrm{equi}}$ flow \\ 
 \hline
 0.2 & 0.187 & 0.150 & 0.146 \\ 
 0.05 & 0.047 & 0.037 & 0.035 \\ 
 \hline
\end{tabular}
\end{center}

We are also interested in the per sample variance of the estimated non-Gaussianity:

\begin{center}
\begin{tabular}{ |c||c|c|c|c| } 
 \hline
  $\tilde{f}_{NL}^{\mathrm{local}}$ training & $\sigma(\tilde{f}_{NL}^{\mathrm{local}})$ training & $\sigma(\tilde{f}_{NL}^{\mathrm{local}})$ flow & $\sigma(\tilde{f}_{NL}^{\mathrm{equi}})$ training & $\sigma(\tilde{f}_{NL}^{\mathrm{equi}})$ flow \\ 
 \hline
 0.2 & 0.074 & 0.085 & 0.044 & 0.040 \\ 
 0.05 & 0.024 & 0.029 & 0.017 & 0.017 \\ 
 \hline
\end{tabular}
\end{center}

We find that the flow is accurate to about $5 \%$ in the mean and to about $10 \%$ in the per sample variance. As discussed in more detail in App.~\ref{sec:nongaussestim} we have normalized the non-Gaussianity estimator from simulations, so that we recover the correct $\tilde{f}_{NL}^{\mathrm{local}}$ value in the training data. We are probing the strongly non-Gaussian regime here, so the variance depends on the mean value, rather than being constant as in the weakly non-Gaussian case.

In this section we have used a correlated prior distribution with a physical power spectrum. When implementing the correlated prior, it is important not to over-count degrees of freedom when evaluating $p(\uvec)$. We are using PyTorch's \texttt{rfftn} function, which uses a slightly degenerate data representation, i.e.\ there is some symmetry in the array of Fourier coefficients. We carefully masked out degenerate information to calculate a correct density value.

\section{Non-Gaussian fields from N-body simulations}
\label{sec:nbodyflow}

Finally we tackle the important use case of representing field PDFs from N-body simuations. We use 100 high-resolution simulations from the publicly available Quijote \cite{Villaescusa-Navarro:2019bje} suite of N-body simulation data to generate training patches of the projected matter field. The simulations were run in a box size of $1 \ h^{-1}\ \text{Gpc}$ and use $1024^3$ DM particles. We use snapshots generated at $z=2$ to estimate the matter density in the 3D simulation volume by painting the dark matter particles on a 3D mesh of size $1024\times1024\times1024$ using the Cloud-in-Cell \cite{1981csup.book.....H} mass assignment scheme as implemented in \texttt{nbodykit} \cite{Hand:2017pqn} library. With this particular choice for the mesh size and redshift, we are able to resolve scales of size $k \gtrsim 1 \ h^{-1}$ Mpc, which are well into the non-linear regime. We subsequently divide the simulation volume into subboxes of size $128\times128\times128$ and project the density along a dimension to get a realization of a 2D field. We do not perform any data augmentation on these patches and generated a total of 51200 independent patches. While for many applications 3-dimensional samples are required, we defer this to future work. 

\begin{figure}[H]
\centering
\includegraphics[width=1.0\textwidth]{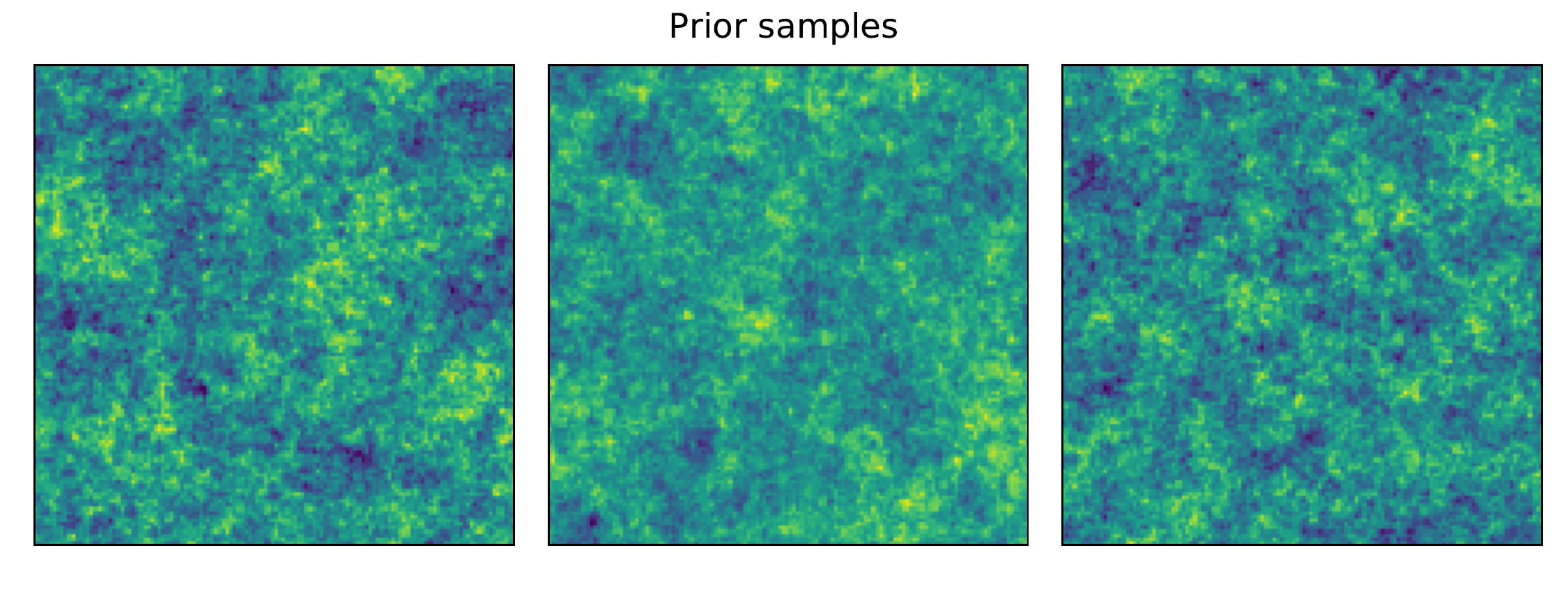}\\
\includegraphics[width=1.0\textwidth]{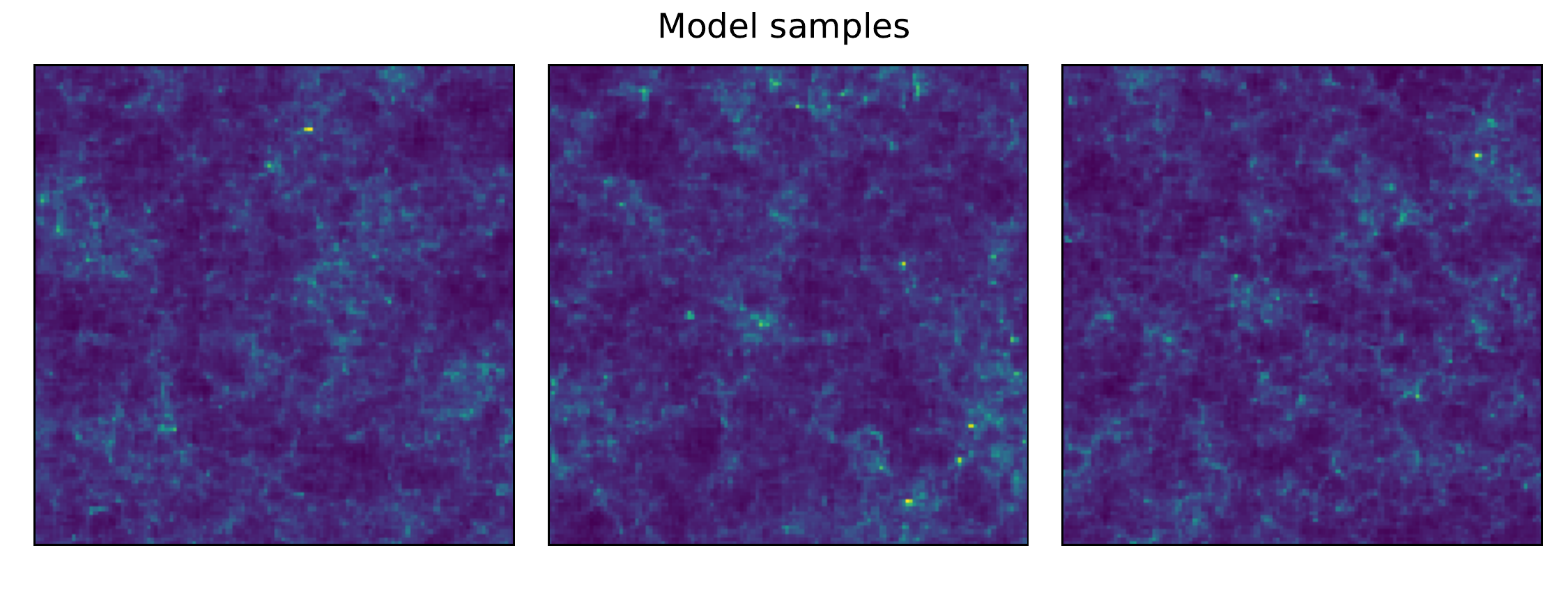}\\
\includegraphics[width=1.0\textwidth]{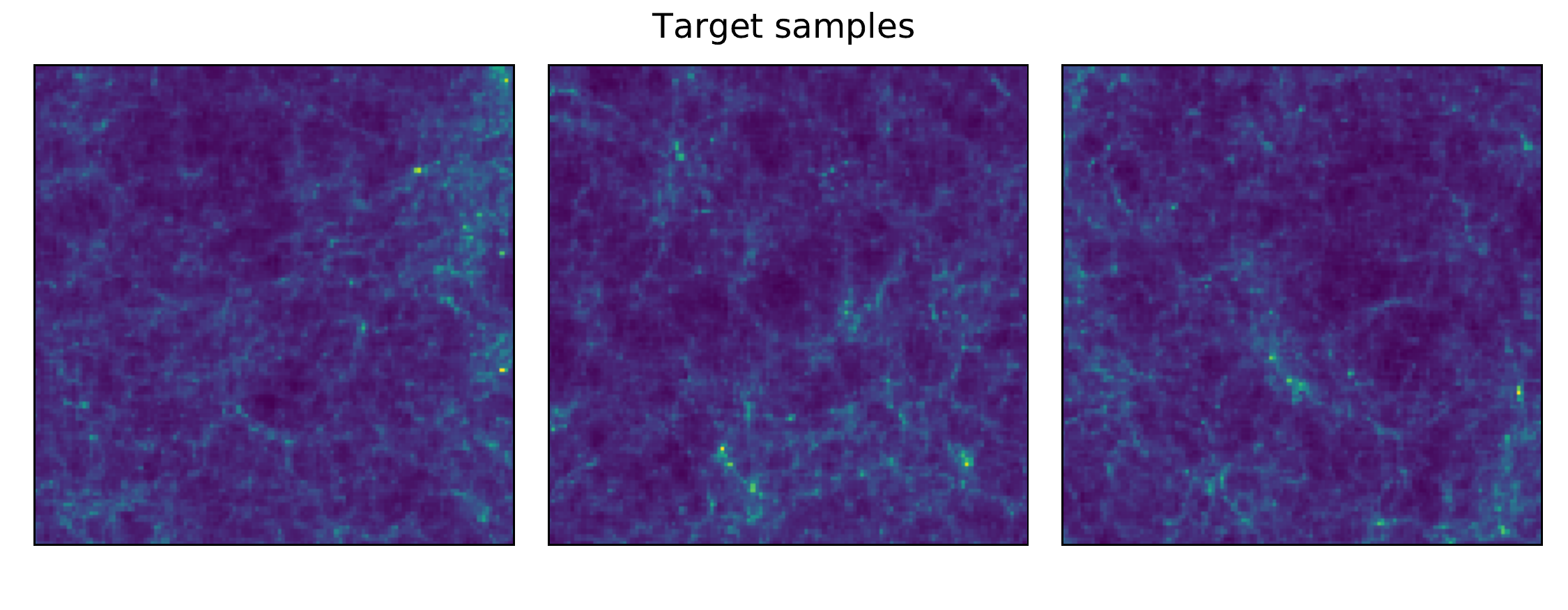}
\caption{N-body simulation example trained on real NVP flow. Each sample is derived from a 3D volume of ($125 \ h^{-1}\ \text{Mpc})^3$ by projecting the matter field along an axis such that the final 2D map has a resolution of $128\times128$ pixels. Top: Samples from the Gaussian prior with the same power spectrum as the samples. Middle: Model samples from the trained flow, derived from the prior samples. It is visible that the flow amplifies the structures seeded in the prior. Bottom: independent samples from the training data.}\label{fig:nbodysamples}
\end{figure}

\begin{figure}[h!]
\centering
\includegraphics[width=0.5\textwidth]{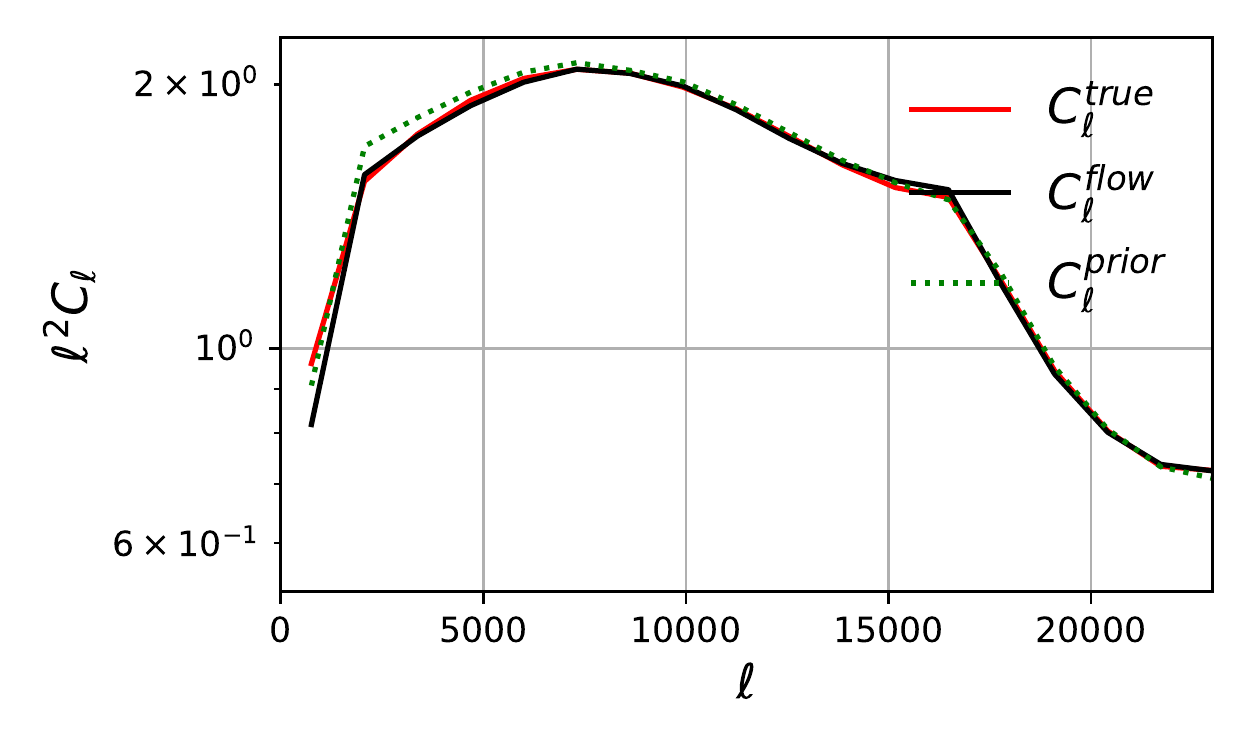}
\caption{Projected matter power spectrum of the N-body simulation samples (red), trained flow model (black) and correlated input prior (green), averaging over 10.000 maps such as those in Fig.~\ref{fig:nbodysamples}. For consistency with previous sections we again use the multipole scale $\ell$ on the x-axis, rather than the comoving wave vector $k$.}\label{fig:nbodyps}
\end{figure}

We trained the same real NVP flow as in the previous section. The training time for this $128\times128$ pixel data set was about 50 hours on our RTX 3090 and used about 11 GB of GPU memory. We found again that it is advantageous to flow from a correlated Gaussian field with the right power spectrum compared to flowing from Gaussian noise. This prior leads to faster convergence and better sample quality. We show random samples from the prior, training and model distribution in Fig.~\ref{fig:nbodysamples}, finding visually excellent results. We note that while there is certainly some correlation between prior and N-body initial conditions on large scales, we are not learning physical structure formation. The prior distribution is only used as a good starting point to flow from.  In Fig.~\ref{fig:nbodyps} we find an almost perfect power spectrum of the flow, with the exception of the smallest $\ell$ bin where the flow power is about $10\%$ too large. We obtain the following non-Gaussianity measurements, sampling over each 10,000 training and model maps:

\begin{center}
\begin{tabular}{ |c|c|c|c| } 
 \hline
 $\tilde{f}_{NL}^{\mathrm{local}}$ training & $\tilde{f}_{NL}^{\mathrm{local}}$ flow & $\tilde{f}_{NL}^{\mathrm{equi}}$ training & $\tilde{f}_{NL}^{\mathrm{equi}}$ flow \\ 
 \hline
 0.85 & 0.78 & 0.61 & 0.62 \\
 \hline
\end{tabular}
\end{center}

\begin{center}
\begin{tabular}{ |c|c|c|c| }
 \hline
   $\sigma(\tilde{f}_{NL}^{\mathrm{local}})$ training & $\sigma(\tilde{f}_{NL}^{\mathrm{local}})$ flow & $\sigma(\tilde{f}_{NL}^{\mathrm{equi}})$ training & $\sigma(\tilde{f}_{NL}^{\mathrm{equi}})$ flow \\ 
 \hline
 0.41 & 0.29 & 0.24 & 0.24 \\ 
 \hline
\end{tabular}
\end{center}

We find that the flow is accurate to $5$ to $10\%$. While this accuracy is sufficient for some applications (such as providing a better prior on non-linear scales than a Gaussian prior), we found that the more complicated Glow flow gives even better results on N-body simulations. We discuss this flow and its training results in App.~\ref{sec:glow}.

The present non-Gaussianity results are not correctly normalized since we did not account for non-Gaussian estimator variance and masking effects. This is not important here since our goal is only to establish equivalence between the training and flow samples, rather than providing physically exact measurements. In fact, the N-body training patches are not periodic (unlike the data in the previous sections), so for both power spectrum and bispectrum estimators we should apodize the mask and numerically estimate the masking bias. Alternatively we could generate periodic training data rather than cutting a larger simulation into patches. For simplicity we have done neither of these two options here. This is justified because our goal is only to compare the N-body and flow results, not to estimate physical parameters. Unlike our real NVP implementation, the Glow flow in App.~\ref{sec:glow} is constructed to be non-periodic.

\section{Conclusion}

The goal of this study was to evaluate how good certain normalizing flows are at representing PDFs of the near-Gaussian random fields that appear in cosmology. Our results are very encouraging. We find that we can train flows whose samples have power spectra and non-Gaussianity accurate to a few percent. Beyond sample generation, we have tested the inverse operation of density evaluation in the analytically tractable Gaussian case and again found percent level accuracy in $\log P$. This opens up the possibility of using flows for inference of cosmological random fields in various ways. 

Unlike typical applications of normalizing flows such as image generation, our physical setup allows us to use a better prior distribution than the commonly used independent Gaussian random variables. We have confirmed that starting with a Gaussian field with a physical power spectrum facilitates learning of the target distribution. This may be crucial for larger maps or 3-dimensional applications. 

We have measured the non-Gaussianity of the flow distribution and compared it to the target distribution, using the bispectrum estimation formalism of cosmology. Exactly measuring three-point correlation functions in statistical machine learning may be useful more generally to qualify the performance of different normalizing flows in representing PDFs.

The main text of the paper was developed with the real NVP flow, which is conceptually simple. However we found in App.~\ref{sec:glow} that in the case of N-body simulations a more complicated architecture, the Glow flow, converges faster and gives higher quality results than the real NVP flow with a correlated prior. Perhaps the Glow architecture could be combined with a correlated prior to be even more efficient.

We have considered relatively small 2-dimensional maps here, up to a side length of 128 pixels. In practice, analysing a small but very deep sky patch with HMC such as for CMB lensing in $\cite{Millea:2020iuw}$ could benefit from a non-Gaussian prior distribution of the lensing potential represented with a normalizing flow. We are in particular interested in applying flows to kinetic Sunyaev-Zeldovich tomography \cite{Deutsch:2017ybc,Smith:2018bpn}, where the scales that generate the kSZ signal are highly non-linear. For this purpose, we have to train a 3-dimensional flow of the matter distribution and develop a method to efficiently patch together many boxes of the small-scale matter distribution into a large-scale CMB map. As discussed in the introduction, the present normalizing flows are also very suitable for variational inference at field level, such as variational inference of the initial conditions of the matter distribution of the universe. Larger maps and 3-dimensional flows will require the use of multi-GPU methods and memory optimized flow architectures. 

For some applications it would be useful to make the flow PDF dependent on astrophysical or cosmological parameters. Flows can depend not only on the parameters of the transformation $T$ but also on learned parameters of the prior distribution. We have not made use of this possiblility here, however in particular the correlated Gaussian prior could easily be extended to depend on parameters that modify the power spectrum. We will explore these directions in upcoming work.

\medskip

\bibliographystyle{unsrt}
\bibliography{main}

\appendix

\section{Non-Gaussianity estimation}
\label{sec:nongaussestim}
In this study, we quantify the performance of the flow samples at the level of the power spectrum (two-point function in momentum space) and of the bispectrum (three-point function in momentum space). In this appendix we recall the standard estimator technology for the three-point function used in cosmology (see e.g.~\cite{Fergusson:2010ia}), adapted to the simple case of a 2D periodic field without sky curvature or mask. 

For a statistically isotropic random field $\phi(\bx)$ its power spectrum is defined by
\begin{align}
\<\phi_{\bk_1}\phi_{\bk_2}\> &= (2\pi)^2 \d_2(\bk_1+\bk_2) P(k_1)\,,
\end{align}
Here $\bk$ are the 2-dimensional wave vectors associated with the random field in Fourier space. Similarily the three-point correlator in momentum space, the so-called bispectrum, $ B(k_1,k_2,k_3)$ is defined by  
\begin{align}
\<\phi_{\bk_1}\phi_{\bk_2}\phi_{\bk_3}\> &= (2\pi)^2 \d_2(\bk_1+\bk_2+\bk_3) B(k_1,k_2,k_3)\,,
\end{align}
where $ \d_2 ({\bf k})$ is the 2-dimensional Dirac $\delta$-function enforcing a triangle condition on the wavevectors ${\bf k}_i$. Due to statistical isotropy, the 3-point function can only be a function of the wavenumbers $k_i = |\bk_i|$ rather than their vector value. Note that our real NVP normalizing flow architecture using periodic padding, CNNs, and a periodic prior guarantees statistical isotropy of the model distribution. 

\textbf{Template estimator.} For the power spectrum, which is a function of a single variable, it is reasonable to simply bin it and measure it in the data as we did above. The bispectrum however is a 3-dimensional quantity and is difficult to visualize and computationally expensive to estimate. Further, in our examples, the bispectrum is smaller than the power spectrum so a plot would mostly show noise. A better method to examine the three point function is thus to fit a set of bispectrum templates to the data and measure their amplitude. Assuming a bispectrum template $B(k_1,k_2,k_3)$, the optimal estimator for its amplitude is (see e.g.~\cite{Babich:2005en}) 

\begin{align}\label{eq:bestimator}
\curl{E} = \frac{1}{6 \ \mathcal{N}} \int \frac{d^2k_1}{(2\pi)^2} \frac{d^2k_2}{(2\pi)^2} \frac{d^2k_3}{(2\pi)^2} \frac{(2\pi)^2 \d_2(\bk_1+\bk_2+\bk_3)B(k_1,k_2,k_3)}{P(k_1)P(k_2)P(k_3)}\phi_{\bk_1}\phi_{\bk_2}\phi_{\bk_3} \,.
\end{align}
This equation can be interpreted as follows: Sum over all combinations of three modes $\bk_i$ that form a triangle, and weight them by their signal-to-noise $\frac{B(k_1,k_2,k_3)}{P(k_1)P(k_2)P(k_3)}$. Here we have assumed that the noise is dominated by the Gaussian contribution, so this estimator is only optimal in the weak non-Gaussian limit. The estimator normalization $\mathcal{N}$ can be calculated analytically in the weak non-Gaussian case, however here we will determine it from simulations to include non-Gaussian variance.

\textbf{Local and equilateral templates.} We are interested in two basic templates for non-Gaussianity, so-called local and equilateral non-Gaussianity. Local non-Gaussianity \cite{Komatsu:2001rj} is generated by transforming a correlated Gaussian field $\phi_{G}(\bx)$ as
\begin{align}
 \phi_{NG}(\bx) = \phi_{G}(\bx) + \tilde{f}_{NL}^{\mathrm{local}}\left( \phi_{G}(\bx)^2 - \< \phi_{G}(\bx)^2 \>\right).
\end{align}
The resulting bispectrum to first order in the non-Gaussianity is 
\begin{align}
 B^{\mathrm{local}}(k_1,k_2,k_3) = \tilde{f}_{NL}^{\mathrm{local}} 2  \big( P_\phi\left(k_1\right) P_\phi(k_2) + P_\phi(k_2) P_\phi(k_3) + P_\phi(k_1) P_\phi\left(k_3\right)\big).
\end{align}
For a scale-invariant power spectrum in 2 dimensions $P(k) = A_s/k^2$ this gives
\begin{align}
 B^{\mathrm{local}}(k_1,k_2,k_3) = \tilde{f}_{NL}^{\mathrm{local}} 2 A_s^2 \left( \frac{1}{k_1^2 k_2^2} + \frac{1}{k_2^2 k_3^2}  + \frac{1}{k_1^2 k_3^2}  \right),
\end{align}
but we will use the former expression in practice. This bispectrum form has the property that it strongly correlates large-scale and small scale modes, i.e.\ $B(k_1,k_2,k_3)$ peaks for so-called squeezed triangles where one $k_i$ is much shorter than the other two. The opposite case is a bispectrum which peaks when all three wave numbers $k_i$ are about the same length. This template is called equilateral non-Gaussianity (e.g.~\cite{Fergusson:2010ia}) and in 2 dimensions we define it as
\begin{align}
 B^{\mathrm{equi}}(k_1,k_2,k_3) = \tilde{f}_{NL}^{\mathrm{equi}} 6 A_s^2 \left( \frac{(k_1+k_2-k_3)(k_2+k_3-k_1)(k_3+k_1-k_2)}{(k_1 k_2 k_3)^{7/3}} \right).
\end{align}
This is an empirical template that is constructed to peak for equilateral configurations (i.e.\ it probes interactions of modes of about the same spatial scales). Unlike for the local bispectrum, map making is not trivial, so we are not generating maps with this specific template here.

\textbf{Efficient position space estimator.} Both of these templates have the property that they can be written in factorizable form, i.e. 
\begin{align}
 B(k_1,k_2,k_3) = A(k_1) B(k_2) C(k_3) + 5 \, \mathrm{ permutations}.
\end{align}
The estimator in Eq.~\ref{eq:bestimator} can then be rewritten in a computationally much more efficient but mathematically equivalent position space form. To do this we insert the identity 
\begin{align}
(2 \pi)^2 \d_2(\bk_1+\bk_2+\bk_3) = \int d^2\bx \, e^{i \bk_1 \bx} \ e^{i \bk_2 \bx} \ e^{i \bk_3 \bx}
\end{align}
and re-organize the terms into a real space integral over the product of Fourier filtered fields. For local non-Gaussianity we obtain 
\begin{align}
    \mathcal{E} = \frac{1}{\mathcal{N}} \int d^2\bx \ A(\bx)^2 \ B(\bx)
\end{align}
with
\begin{align}
    A(\bx) = \int \frac{d^2k}{(2\pi)^2} \phi(k)  e^{i \bk \bx}, \hspace{2cm}
    B(\bx) = \int \frac{d^2k}{(2\pi)^2} \frac{\phi(k)}{P_\phi(k)}  e^{i \bk \bx}. 
\end{align}
The expression for equilateral non-Gaussianity is very similar:
\begin{align}
    \mathcal{E} = \frac{1}{\mathcal{N}} \int d^2\bx \ A_s^{-1}\left[ -3 \ \tilde{A}(\bx) \ \tilde{B}^2(\bx) + 6 \tilde{B}(\bx) \tilde{C}(\bx) \tilde{D}(\bx) -2 \tilde{D}^3(\bx) \right]
\end{align}
with 
\begin{align}
    \tilde{A}(\bx) = \int \frac{d^2k}{(2\pi)^2} k^{8/3} \phi(k)  e^{i \bk \bx}, \hspace{2cm}
    \tilde{B}(\bx) = \int \frac{d^2k}{(2\pi)^2} k^{-1/3} \phi(k)  e^{i \bk \bx}, \\
    \tilde{C}(\bx) = \int \frac{d^2k}{(2\pi)^2} k^{5/3} \phi(k)  e^{i \bk \bx}, \hspace{2cm}
    \tilde{D}(\bx) = \int \frac{d^2k}{(2\pi)^2} k^{2/3} \phi(k)  e^{i \bk \bx}. 
\end{align}
Note that these expressions are for scale-invariant non-Gaussianity in two dimensions, and thus slightly different from their three-dimensional counterparts more commonly used in cosmology.

\section{The Glow flow applied to N-body simulations}
\label{sec:glow}

A well known modification and generalisation of the real NVP flow is the Glow flow \cite{2018arXiv180703039K}. Glow has been shown to generate better RGB images than real NVP, and we thus examine if it also improves results on cosmological data. On the other hand, the flow architecture is more complicated, in particular it does not flow from a spatially organized prior, so we cannot straightforwardly flow from a correlated Gaussian prior as we did for real NVP. It would also be more difficult to enforce periodicity, which however for the present application is not required.

\subsection{Flow architecture}
Glow uses a multi-scale architecture with a series of $K$ steps of flow, each step consisting of an activation normalization (or \textit{actnorm}) layer, an invertible $1\times 1$ convolution, and finally an affine coupling layer. The actnorm layer, introduced in \cite{2018arXiv180703039K}, is a generalisation of batch normalization, initializing trainable scale and bias parameters on each batch such that \textit{each channel} has zero mean and unit variance. The learnable $1\times1$ convolution mixes channels, a generalisation of permuting the channels or alternating the subsets $x_1$ and $x_2$ in the checkerboard masking as in real NVP. The affine coupling layers have learned functions $s$ and $t$ as described in Sec.~2.1. We use 3 convolutional layers with kernel sizes 3, then 1, then 3. The two hidden layers have ReLU activation functions and 128 channels each. The final convolution of each CNN is initialized with zeros as recommended in \cite{2018arXiv180703039K}.

\begin{figure}[!h]
\centering
\includegraphics[width=0.325\textwidth]{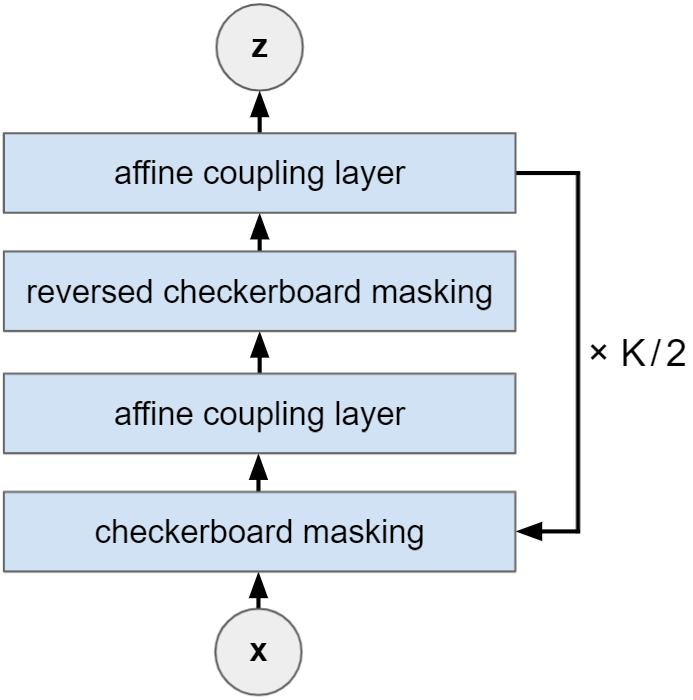}
\quad
\quad
\quad
\includegraphics[width=0.6\textwidth]{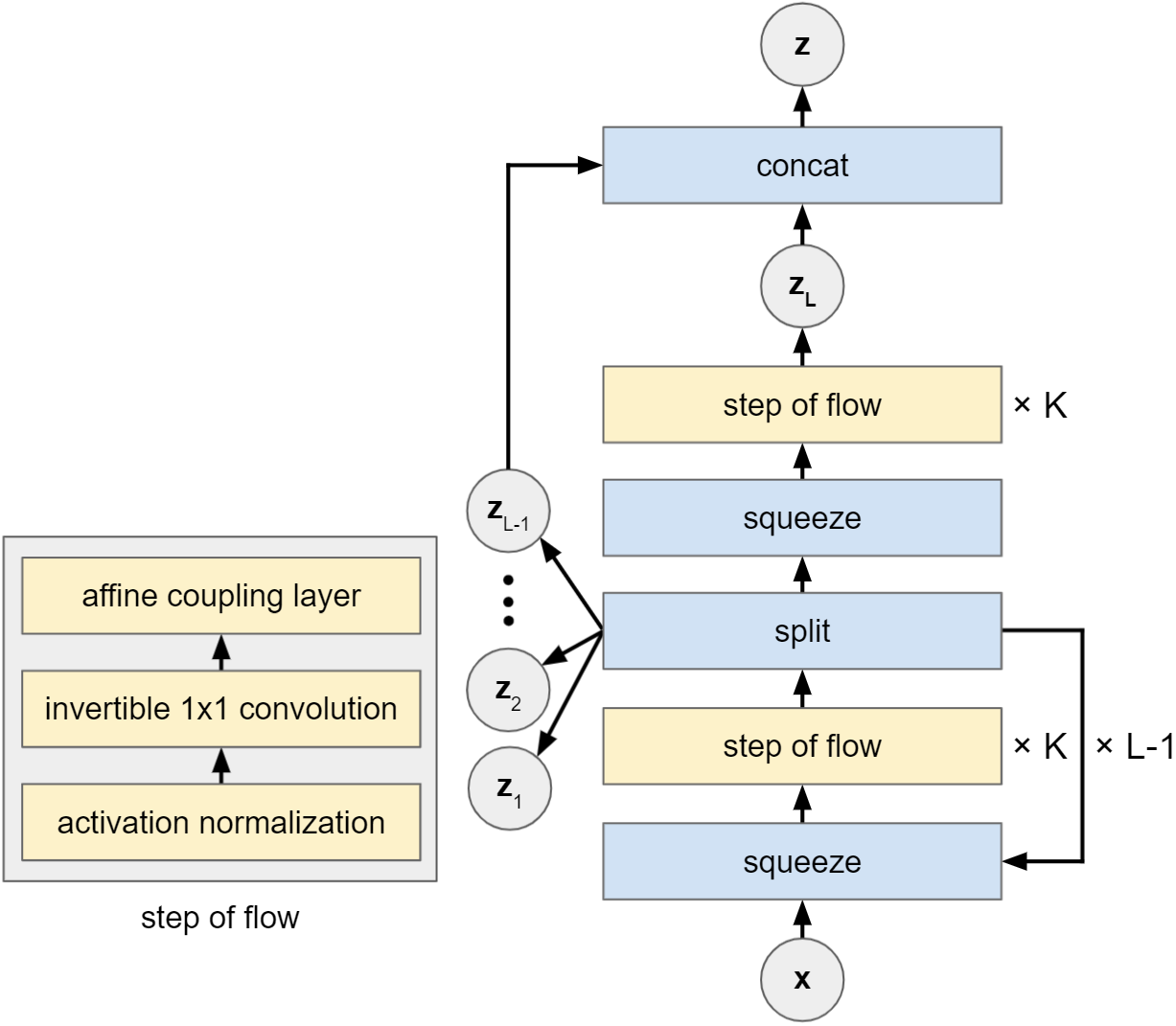}
\subfloat[\label{affine_arch} Real NVP architecture]{\hspace{0.275\linewidth}}
\subfloat[\label{glow_arch} Glow architecture]{\hspace{0.8\linewidth}}

\caption{Comparison of our real NVP architecture and the Glow architecture. The multi-scale architecture of Glow improves training accuracy at small $l$.}\label{fig:architectures}
\end{figure}

Prior to the $K$ steps of flow, a squeeze operation reshapes each $2\times2\times c$ region of the input tensor to $1\times1\times 4c$. The squeeze operation thus reshapes the input tensors from $n\times n\times c$ to $\frac{n}{2}\times\frac{n}{2}\times4c$. After $K$ steps of flow, the tensor is split into halves along the channel dimension. The squeeze, flow steps, and split form an $L-1$ number of levels, followed by a final level not containing a split function. A comparison of the Glow architecture with the affine coupling layer and checkerboard masking flow used in the body of this paper is shown in Fig.~\ref{fig:architectures}. A further difference is that our implementation of Glow, unlike our real NVP flow, is not periodic. This is suitable for the present N-body training data.

We use a PyTorch implementation of Glow taken from \cite{GlowGithub} that has been slightly modified to include the loss function Eq.~\ref{eq:loss} and use the Adam optimizer.

\subsection{Application to N-body simulations}

\begin{figure}[h!]
\centering
\includegraphics[width = 0.45\textwidth]{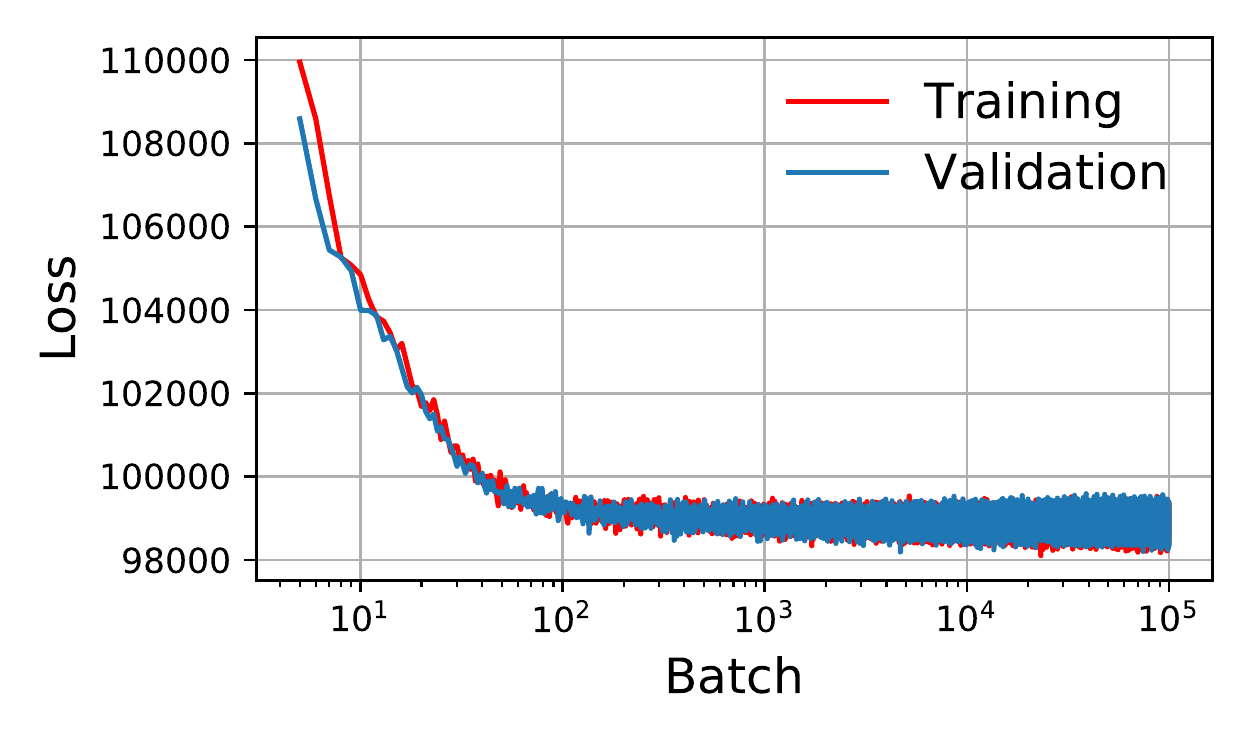}
\includegraphics[width=0.45\textwidth]{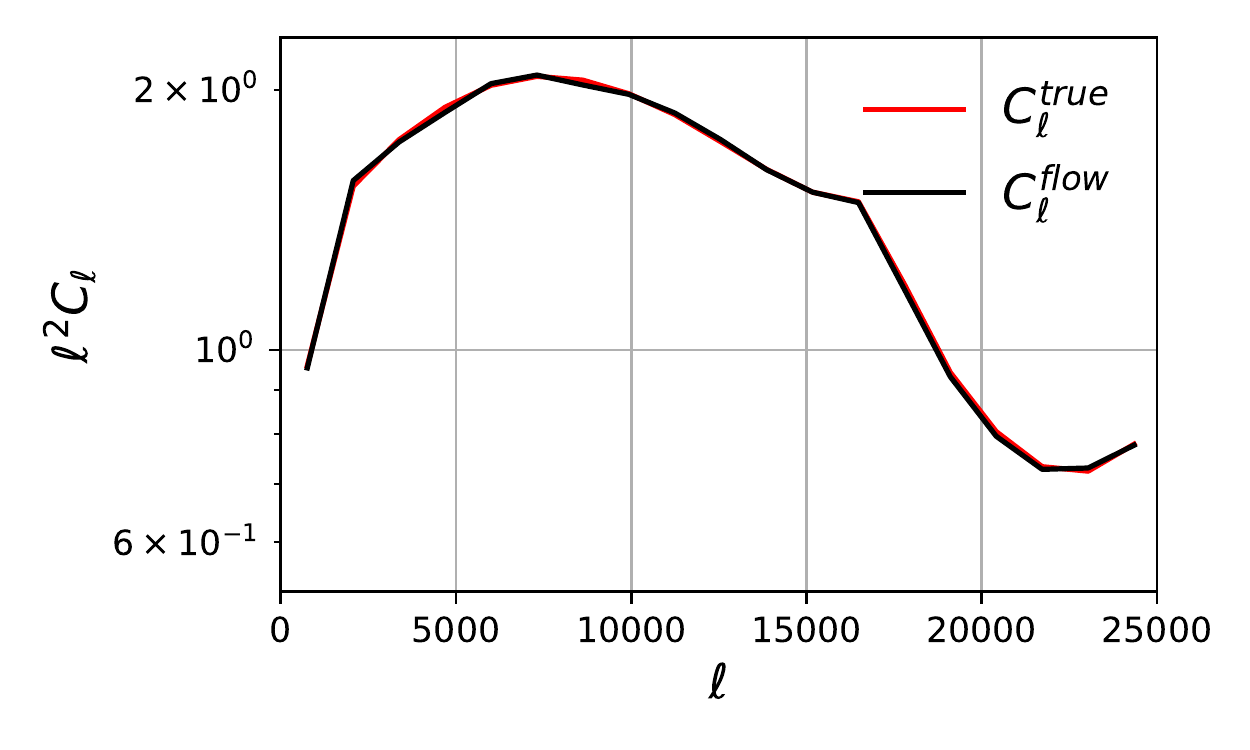}\\
\caption{Left: Loss during training with Glow (starting from batch 5). The total training time in this plot is about 20 hours on an RTX 3090, but the flow is trained out after about 2 hours. Unlike in previous sections, the loss falls so quickly that we plot it on a log scale. The variance of the loss is not growing (this is an effect of the point density in the plot). After about 10,000 batches we enter a regime of mild over-training and we stop the training. Right: The power spectrum of the training data and of the flow samples agrees perfectly.}\label{fig:glow_nbody_ps}
\end{figure}

\begin{figure}[h!]
\centering
\includegraphics[width=1.0\textwidth]{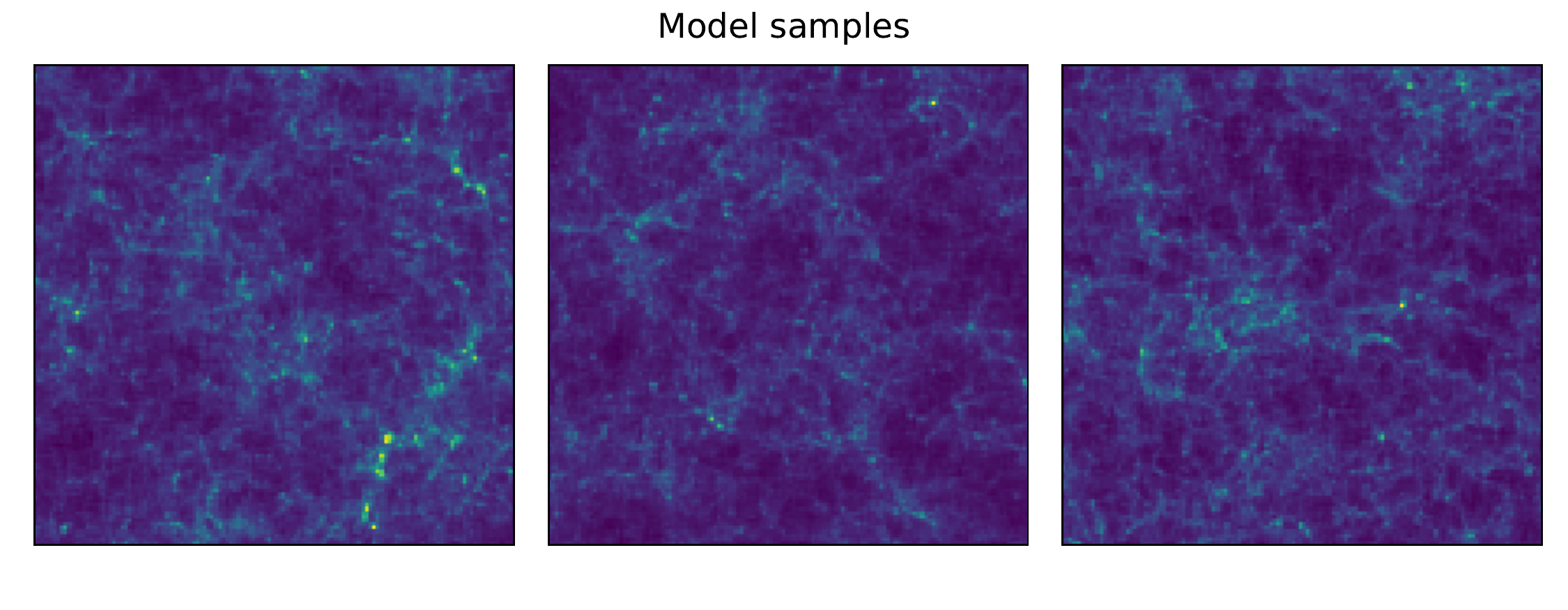}\\
\includegraphics[width=1.0\textwidth]{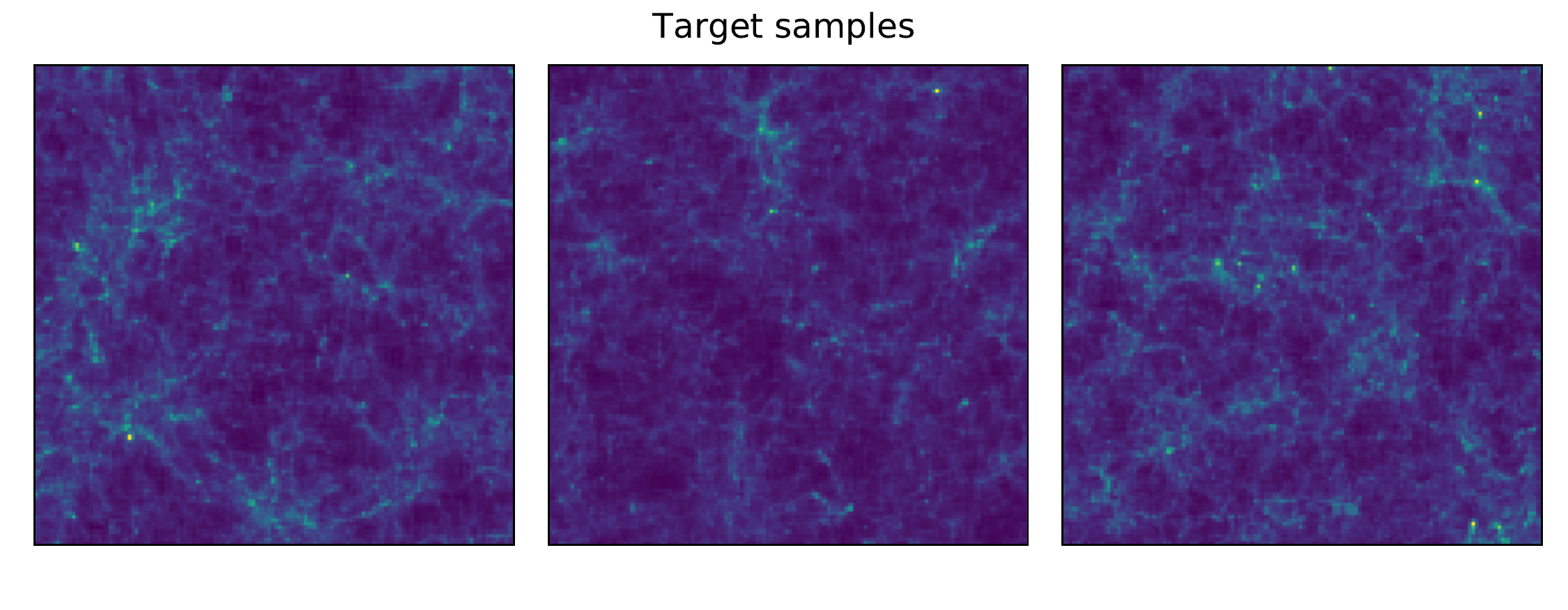}
\caption{N-body simulation example from Sec.~\ref{sec:nbodyflow} trained with the Glow flow. Top: Model samples from the trained flow. Bottom: Samples from the training data. Training and model samples cannot be distinguished visually.}\label{fig:nbodysamples_glow}
\end{figure}

The performance of Glow on cosmological data is demonstrated here by training on the N-body patches introduced in Sec.~5. We train with a depth of flow $K=16$ and $L=3$ levels; this provides a total of 1,592,704 trainable parameters. We use Adam optimization with a learning rate of $5\times10^{-4}$ and batch size of 32, using about 11 GB of GPU memory. The training converges extremely quickly as illustrated in Fig.~\ref{fig:glow_nbody_ps} (left). Samples from the N-body training are shown in Fig.~\ref{fig:nbodysamples_glow}, and the power spectra in Fig.~\ref{fig:glow_nbody_ps} (right). The Glow flow has a near perfect power spectrum, including in the smallest $l$ bin (largest spatial scale) where the basic checkerboard real NVP flow in Sec.~5 was suboptimal. We obtain the following non-Gaussianity measurements, averaging over 10,000 training and model samples:

\begin{center}
\begin{tabular}{ |c|c|c|c| } 
 \hline
 $\tilde{f}_{NL}^{\mathrm{local}}$ training & $\tilde{f}_{NL}^{\mathrm{local}}$ flow & $\tilde{f}_{NL}^{\mathrm{equi}}$ training & $\tilde{f}_{NL}^{\mathrm{equi}}$ flow \\ 
 \hline
 0.85 & 0.85 & 0.61 & 0.63 \\
 \hline
\end{tabular}
\end{center}

\begin{center}
\begin{tabular}{ |c|c|c|c| } 
 \hline
   $\sigma(\tilde{f}_{NL}^{\mathrm{local}})$ training & $\sigma(\tilde{f}_{NL}^{\mathrm{local}})$ flow & $\sigma(\tilde{f}_{NL}^{\mathrm{equi}})$ training & $\sigma(\tilde{f}_{NL}^{\mathrm{equi}})$ flow \\ 
 \hline
 0.40 & 0.37 & 0.24 & 0.30 \\ 
 \hline
\end{tabular}
\end{center}

We find that the Glow flow is accurate at the percent level in $\tilde{f}_{NL}^{\mathrm{local}}$, an improvement over the real NVP flow in Sec.~\ref{sec:nbodyflow}, and has similar performance in $\tilde{f}_{NL}^{\mathrm{equi}}$. As in Sec.~\ref{sec:nbodyflow} we have not corrected for the masking bias as our goal was only to show equivalence between training and model distribution.

While results presented here used 11 GB of memory for 128 channels in the two hidden layers, we find almost the same performance reducing to 32 channels in each hidden layer, which reduces the GPU memory size to 4.1 GB. The 32 channel version has a significant error only in the highest $l$ bin, being about 10\% in the power spectrum. When extending to 3-dimensional data, using a memory efficient architecture will be important.

\end{document}

%% file: math_commands.tex
\newcommand{\fnl}{{f_\mathrm{NL}}}

\newcommand{\uvec}{{\bf u}}
\newcommand{\xvec}{{\bf x}}

\usepackage{amsmath,amsfonts,bm}

\def\eqref#1{equation~\ref{#1}}

\def\1{\bm{1}}

\def\rvu{{\mathbf{i}}}

\def\rvu{{\mathbf{u}}}

\def\rvx{{\mathbf{x}}}

\def\rvz{{\mathbf{z}}}

\DeclareMathAlphabet{\mathsfit}{\encodingdefault}{\sfdefault}{m}{sl}
\SetMathAlphabet{\mathsfit}{bold}{\encodingdefault}{\sfdefault}{bx}{n}

\newcommand{\abs}[1]{\left|#1\right|}

\newcommand{\createDensityWithSub}[2]{{#1}_{\mathrm{#2}}}

\newcommand{\pu}{\createDensityWithSub{p}{u}}

\def\d{\delta}

\def\={\nonumber &=}

\def\&{{}&}

\def\({\left(}
\def\){\right)}
\def\[{\left[}
\def\]{\right]}
\def\<{\left\langle}
\def\>{\right\rangle}

\def\bk{{\bf k}}

\def\bx{{\bf x}}

\def\curl{\mathcal}

\def\eq{\begin{align}}
\def\qe{\end{align}}
\def\eqa{\begin{eqnarray}}
\def\qea{\end{eqnarray}}

\def\and{\quad \mbox{and} \quad}

\def\fnl{f_\textrm{NL}}

\def\bfnl{\kern2pt\overline{\kern-2ptf}_\textrm{NL}}

\def\barQ{\kern2pt\overline{\kern-2pt\curl{Q}}}

\def\barR{\kern2pt\overline{\kern-2pt\curl{R}}}
